\documentclass[aps,pra,reprint,superscriptaddress,nofootinbib]{revtex4-2}

\usepackage{graphicx}
\usepackage{amsmath}

\usepackage{siunitx}
\usepackage{braket}
\usepackage{bbold}
\usepackage{booktabs}

\usepackage{hyperref}
\usepackage[capitalize,nameinlink]{cleveref}
\DeclareMathAlphabet{\stdmathbb}{U}{msb}{m}{n}

\makeatletter
\DeclareRobustCommand\ket[1]{%
  \@ifnextchar\bra{\k@t{#1}\!}{\k@t{#1}}%
}
\newcommand\k@t[1]{{|{#1}\rangle}}
\makeatother

\renewcommand{\l}[1]{\Ket{\mathbb{#1}}}
\newcommand{\el}[1]{\Ket{\mathcal{E}_1 \mathbb{#1}}}
\newcommand{\eel}[1]{\Ket{\mathcal{E}_2 \mathbb{#1}}}

\newcommand{\jz}{\hat J_z}
\newcommand{\jx}{\hat J_x}
\newcommand{\jy}{\hat J_y}

\crefname{extendedfigure}{Extended Data Fig.}{Extended Data Figs.}
\Crefname{extendedfigure}{Extended Data Fig.}{Extended Data Figs.}
\crefname{supplementfigure}{Supplementary Fig.}{Supplementary Figs.}
\Crefname{supplementfigure}{Supplementary Fig.}{Supplementary Figs.}
\crefname{supplementtable}{Supplementary Table}{Supplementary Tables}
\Crefname{supplementtable}{Supplementary Table}{Supplementary Tables}

\begin{document}

\title{Error-tolerant gate set for single-ion logical qubits}

\author{Kyle DeBry}
\email{debry@mit.edu}
\affiliation{Massachusetts Institute of Technology, Cambridge, MA}
\affiliation{Lincoln Laboratory, Massachusetts Institute of Technology, Lexington, MA}
\author{Agustin Valdes Martinez}
\altaffiliation{Present address: University of Chicago, Chicago, IL}
\affiliation{Massachusetts Institute of Technology, Cambridge, MA}
\affiliation{Lincoln Laboratory, Massachusetts Institute of Technology, Lexington, MA}
\author{Alexandra Plocki}
\affiliation{Massachusetts Institute of Technology, Cambridge, MA}
\affiliation{Lincoln Laboratory, Massachusetts Institute of Technology, Lexington, MA}
\author{Gabriel J. Gregory}
\affiliation{University of Oregon, Eugene, OR}
\author{Timothy J. Burke}
\affiliation{Massachusetts Institute of Technology, Cambridge, MA}
\author{Xiaoyang Shi}
\affiliation{Massachusetts Institute of Technology, Cambridge, MA}
\author{Ethan R. Clements}
\affiliation{Lincoln Laboratory, Massachusetts Institute of Technology, Lexington, MA}
\author{David Reens}
\affiliation{Lincoln Laboratory, Massachusetts Institute of Technology, Lexington, MA}
\author{William J. Setzer}
\affiliation{Lincoln Laboratory, Massachusetts Institute of Technology, Lexington, MA}
\author{May E. Kim}
\affiliation{Lincoln Laboratory, Massachusetts Institute of Technology, Lexington, MA}
\author{Jameson O'Reilly}
\affiliation{University of Oregon, Eugene, OR}
\author{John Chiaverini}
\altaffiliation{Present address: IonQ, Inc., Boston, MA}
\affiliation{Massachusetts Institute of Technology, Cambridge, MA}
\affiliation{Lincoln Laboratory, Massachusetts Institute of Technology, Lexington, MA}
\author{Isaac L. Chuang}
\affiliation{Massachusetts Institute of Technology, Cambridge, MA}
\author{Colin D. Bruzewicz}
\affiliation{Lincoln Laboratory, Massachusetts Institute of Technology, Lexington, MA}
\author{Robert McConnell}
\affiliation{Lincoln Laboratory, Massachusetts Institute of Technology, Lexington, MA}

\date{\today}

\begin{abstract}
Quantum error correction protects quantum information from noise, but typically requires many physical qubits to encode each logical qubit, linked by entangling gates that are often the noisiest operations in a processor \cite{Bal14,Mos23,Fow08}. Encoding a logical qubit in the many internal states of a single atom \cite{Gro20} reduces this overhead and has protected quantum memories beyond break-even \cite{Deb25,Li25d}. Entangling gates, however, act via shared degrees of freedom and produce correlated errors, and no single-atom code has yet protected against them. Here we demonstrate a gate set for spin-cat logical qubits \cite{Yu25,Oma24b,Lim24} encoded in single trapped ions, with an entangling gate designed such that its own errors are detectable or correctable by the code. The gate couples each ion to the shared motional mode via the same internal-state operator the code protects from, so motional errors appear as powers of that operator on each ion and can be handled locally, without additional ions. Conditional on no detected erasure, we measure a logical entangling-gate infidelity of $1.9(1.3)\times 10^{-5}$, in agreement with an independent error budget, flagging $99.97\%$ of all errors as erasures. We also realize single-qubit Cliffords with an undetected error of $1.5\times 10^{-5}$ and a logical $T$ gate, and show that local error correction reduces gate errors under injected noise. Protecting logical computation, not only memory, within individual atoms opens a route to fault-tolerant quantum computing with substantially fewer atoms.
\end{abstract}

\maketitle

\section{Introduction}

\begin{figure*}
    \includegraphics[width=\linewidth]{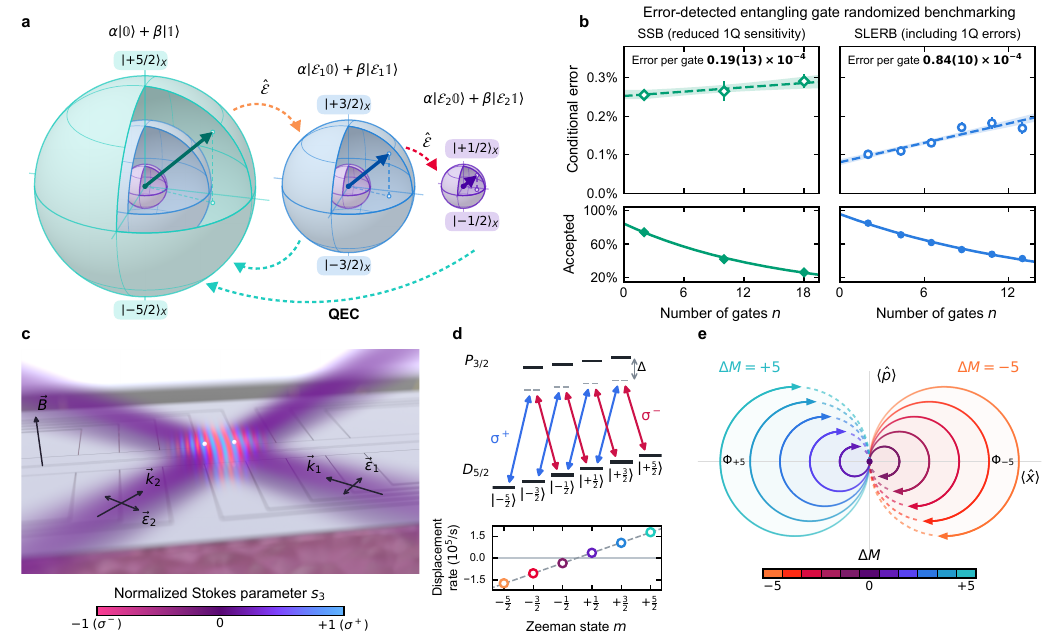}
        \centering
    \caption{\textbf{Logical encoding and multi-level, error-correctable entangling gate.}
    \textbf{a,} Visualization of the spin-cat encoding in the $D_{5/2}$ manifold of $^{40}\text{Ca}^+$ as three concentric Bloch spheres, where the logical amplitudes $\alpha$ and $\beta$ describing the Bloch vector's direction are preserved by errors. Teal, blue, and purple spheres represent the logical, first-order, and second-order error subspaces, respectively. Dashed arrows denote errors and QEC operations changing subspace but preserving the logical information. 
    \textbf{b,} Randomized benchmarking using subspace leakage error randomized benchmarking (SLERB) and symmetric stabilizer benchmarking (SSB), showing fits with $1\sigma$ confidence regions. Accepted fraction is the fraction of trials that had no detected erasures. SLERB sequences take data as a function of the number of Cliffords, which include a variable number of entangling gates, so the points are plotted at the average number of gates. 1Q: single qubit.
    \textbf{c,} Illustration (not to scale) of two ions on a surface-electrode trap illuminated by the two Raman beams, showing the polarization running wave's normalized Stokes parameter $s_3$. The magnetic field $\vec{B}$ is out of the plane of the trap.
    \textbf{d,} Depiction of one ion's six-level internal state coupled to the continuous-variable motional state, used to implement the entangling interaction. On top is an energy-level diagram showing the off-resonant Raman beams detuned by $\Delta$ from the $P_{3/2}$ states, and below is the measured rate of displacement of the coherent state of one ion under this force as a function of $m$. Error bars show one standard error but are smaller than the markers, and the line is a linear fit through the origin. 
    \textbf{e,} Idealized trajectories of the ions' shared motional phase space for collective spin difference $\Delta M=m_1-m_2$. Enclosed areas determine $\Delta M$-dependent geometric phase $\Phi_{\Delta M}$.}
    \label{fig:logical-overview}

\end{figure*}
Large-scale quantum computation will require quantum error correction, which protects logical information from errors as long as the physical error rate is below the threshold of the code being used \cite{nielsen_quantum_2010,Sho95,Aha96}. The number of physical qubits and operations this requires depends strongly on how far below threshold the hardware operates \cite{Fow08}, and in most platforms it is the entangling gate that sets this margin \cite{Gae16,Mos23,Eve23,Eve26}. The most promising applications of quantum computing remain out of reach on current hardware, despite substantial improvements to entangling-gate fidelities and physical qubit counts over the past two decades \cite{Bru19,Kja19,Hen20}, in part because of the large physical-qubit overheads required by current codes of interest \cite{Cam16,Bev22}.

A potential route to fault tolerance that can mitigate the large physical-qubit overhead is to carry out part of the error correction within each constituent physical system; atoms are natural candidates for such encodings. Atoms have many internal levels that can be controlled with high fidelity \cite{Lim23,And14,Rin21,Low23,Shi26,Deb23}, and multilevel encodings such as spin-cat codes use this larger Hilbert space to create a logical qubit in a single atom \cite{Gro20,Oma24b,Lim24,Ong26,Roy25,Yu25}. These encodings have recently been shown to protect stored quantum information beyond break-even \cite{Deb25,Li25d}. That protection, however, is obtained only for memory errors, which occur while the qubit is idle. Although gate errors have been converted to erasures (errors with known locations) in single neutral atoms \cite{Ma23,Sch23,Eve26} and ions \cite{Qui24}, and bosonic encodings have made rapid progress \cite{Mat24b,Fon26,Ni26}, no single-atom code has yet been used to correct entangling-gate errors.

Here we develop a logical gate set that extends spin-cat codes to also protect entangling operations. The gate set includes universal single-ion operations via Clifford gates and a $T$ gate, as well as a multi-level entangling gate. The spin-$5/2$ spin-cat code (with dimension $d=6$) can be pictured as a set of three concentric Bloch spheres, with the logical qubit on the outer shell and error states on inner shells reached by successive powers of $\jz$ errors as shown in \cref{fig:logical-overview}a. The operators $\jx$, $\jy$, and $\jz$ are the $x$, $y$, and $z$ projections of total angular momentum, respectively.
An error moves the state from one shell to another without changing its logical orientation, which results in protection from idling errors \cite{Deb25,Chi20,Lim24}.
In order to retain this protection during computation, our error-tolerant single-qubit gate set performs the same unitary operation on each of the three shells, preserving the logical information with or without an error \cite{Kap17,Ma19b,Ong26}. To protect against errors in the gate operation itself, we engineer the entangling gate such that its dominant errors are detectable and correctable using the spin-cat codewords.

The gate set's single-qubit logical operations are implemented via arbitrary $\mathrm{SU}(6)$ control of the $D_{5/2}$ manifold, while the entangling gate is implemented via a multi-level spin-dependent force. Radiofrequency (RF) fields perform $\mathrm{SU}(2)$ rotations of the spin manifolds about $\jx$ and $\jy$, where $\jx$, $\jy$,  is the projection of angular momentum along the $\hat{\epsilon}$ direction. A tensor light shift performs a non-$\mathrm{SU}(2)$ $\jz^2$ rotation of the manifold, which is combined with the RF operations to create the Clifford gates and $T$ gate. The entangling gate is implemented using two far-detuned Raman lasers that create a force on the ions proportional to $\jz$, producing geometric phases that entangle the logical qubits.

With this construction, the entangling gate is dominated by errors that the spin-cat code can correct locally.
In trapped ions, entangling-gate errors often come from motional-mode effects such as those due to noise on the applied force's intensity and phase, and normal-mode heating \cite{Win97,Sut22b}, and they are generally correlated between the two ions. 
Because our ion--motion coupling is linear in $\jz$, motional errors also manifest as powers of $\jz$ such that they commute with the action of the gate and are protected by the code.
Gate errors now decompose into products of powers of $\jz$ on each ion, and can thus be corrected or detected on each ion individually (using no additional entangling operations), moving each ion back to the logical (outer) shell.

Using these codes, we demonstrate both correction and detection of entangling-gate errors and single-qubit idling errors between gates.
Using error detection, we measure a logical error per entangling gate, conditioned on not detecting an erasure, of $0.19(13)\times 10^{-4}$ (one-sided 95\% upper bound $4\times10^{-5}$), in agreement with an independently calibrated error budget of $2.3(2)\times10^{-5}$.
We also demonstrate error correction under injected Raman intensity errors and motional kicks (\cref{fig:logical-gate}b,c), with local error correction improving fidelity for sufficiently strong noise strengths.
Logical single-qubit operations can also be performed with high fidelity, and we measure a logical error of $1.5(3)\times10^{-5}$ per logical single-qubit Clifford gate for a detection-only Clifford gate set.
While these results rely on excluding detected errors, these errors can be treated as erasures. A suitably designed code can tolerate erasures at much higher rates than generic errors so that fewer atoms and fewer operations are required to perform the same task \cite{Wu22,Kan22,baranes2026leveraging,Per24b,Zha25j}.
Based on these results, we propose an error-detected logical quantum charge-coupled device (QCCD) architecture \cite{Win97,Pin20} in which each single-ion logical qubit can reach an exceptionally low undetected error rate, where the resulting erasures can be corrected efficiently by an outer code such as an erasure-optimized surface code.

\section{Error-tolerant gate set}

\subsection{Encoding and error tolerance}
We achieve protection from $\jz$ errors, which describe dephasing of the ion's internal states, by encoding a spin-cat logical qubit in the $D_{5/2}$ ($J=5/2$) manifold of a $^{40}\text{Ca}^+$ ion trapped in a cryogenic surface-electrode trap (Methods). The logical codewords are the extremal eigenstates of $\jx$, $\l0=\Ket{-5/2}_X$ and $\l1=\Ket{+5/2}_X$, where $\Ket{m}_X$ is the eigenstate of $\jx$ with eigenvalue $m$. The codewords are given in the $\jz$ basis in \cref{eq:logical-codewords-0,eq:logical-codewords-1} (Methods; \cref{fig:methods-codewords}a). The codewords are prepared by optically pumping the ion into the $S_{1/2}$ ground state's $m=-1/2$ level, applying a spectrally selective $\pi$ pulse on the \SI{729}{\nano\meter} quadrupole transition to the $\Ket{-5/2}$ state of the $D_{5/2}$ manifold, and then applying a radiofrequency (RF) $\pi/2$ pulse to rotate the state into $\l0$ or $\l1$ (Methods).

Dephasing errors acting on these codewords move the state between shells (\cref{fig:logical-overview}a) without altering the logical information. The errors can be described by random rotations $e^{-i\jz \phi_z}$ where $\phi_z$ is a fluctuating rotation angle, which result in the decay of the coherences of a density matrix. These continuous errors can be discretized into a power series of error operators $\hat{\mathcal{E}}_k=\jz^k$ when $\lvert\phi_z\rvert \ll 1$ \cite{Chi20,Deb25}. A first-order error $\jz$ shifts the $\jx$ eigenvalue by one unit and acts with equal strength on both codewords, so it maps the logical shell onto a first error shell spanned by $\el0=\Ket{-3/2}_X$ and $\el1=\Ket{+3/2}_X$ without changing the logical amplitudes (\cref{fig:logical-overview}a). A second application of $\jz$ has a component on the second error shell, $\eel0=\Ket{-1/2}_X$ and $\eel1=\Ket{+1/2}_X$, completing the three concentric Bloch spheres of \cref{fig:logical-overview}a. Because the two codewords differ in their $\jx$ projection by five units of angular momentum, errors up to $\jz^2$ can be corrected and errors up to $\jz^4$ can be detected (Methods). If an error process that is first order in $\jz$ (e.g. motional heating during the gate) leads to raw Bell infidelity $p \ll 1$, the error probability is reduced to $p^3$ by ideal correction and $p^5$ by detection.

For a logical gate to preserve this protection it must act equivalently on all three shells (up to an overall phase on each shell), with no coupling between them \cite{Kap17,Ma19b,Gup24}. A logical gate $\hat U_L$ is therefore implemented as the direct sum $\hat U_0\oplus\hat U_1\oplus\hat U_2$, where $\hat U_i$ is the same logical unitary acting on the $i$-th shell. A correctable error that occurs before such a gate is still correctable after it. The entangling interaction satisfies a stronger condition: it is diagonal in the $\jz$ basis and commutes with $\jz$ during the spin-dependent force pulses, so errors that occur during these pulses are also correctable. The logical entangling gate includes a spin-echo $\jy$ rotation at its midpoint, which is susceptible to detuning errors. The single-qubit gates, by contrast, act identically on all three shells only as complete operations, so errors between them are correctable but errors partway through a gate generally are not (Methods). Thus, for errors occurring at any time except during the application of single-qubit operations (including the entangling gate's spin-echo pulse), the set of correctable errors is $\operatorname{span}\{\jz^k \mid k =0,1,2\}$, and the set of detectable errors $\operatorname{span}\{\jz^k \mid k=0,\ldots,4\}$.

\begin{figure*}
    \includegraphics[width=1\linewidth]{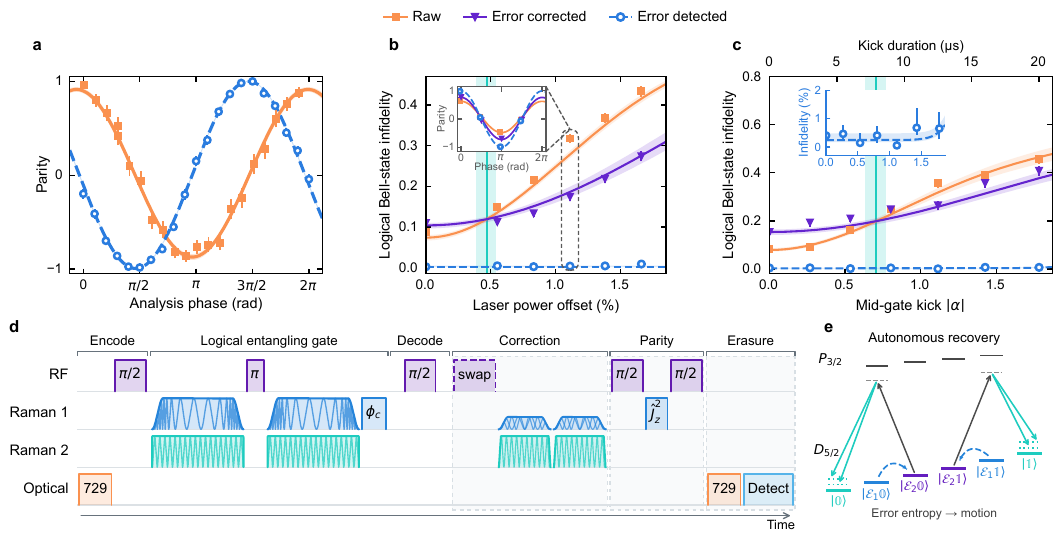}
    \caption{\textbf{Logical entangling gate benchmarking and correction of motional errors.}
    \textbf{a,} Parity oscillations of the logical Bell state, raw and with error detection. Lines are fits and shaded bands are $1\sigma$ fit uncertainties.
    \textbf{b,} Experimental results for logical Bell-state error versus injected Raman power offset for unmitigated, error-corrected, and error-detected gates, demonstrating decreased error from both error detection and correction. Lines and shaded bands show predictions from independent measurements with $1\sigma$ uncertainties, and error bars denote $1\sigma$ fidelity uncertainties. Vertical line and shaded region show estimated break-even power offset value and its uncertainty. Inset: the three parity scans taken at a 1.11\% power offset.
    \textbf{c,} Logical Bell-state infidelity versus the magnitude $|\alpha|$ of a spin-independent displacement with random phase applied midway through the first loop of the gate (top axis, corresponding kick duration), raw, with recovery (corrected) and with error detection. Vertical line and shaded region show estimated break-even mid-gate kick error value and its uncertainty. Inset: infidelity with error detection on an expanded scale. Lines are predictions from independent calibrations, with $1\sigma$ bands. Recovery lowers the error for $|\alpha| > 0.7(1)$. 
    \textbf{d,} Pulse sequence for encoding, entangling, decoding, two-mode error correction, parity analysis, erasure detection and readout; optional stages are shown in dashed boxes.
    \textbf{e,} Autonomous recovery sequence used for error correction of logical Bell states. First, if necessary (\textbf{c}), the error populations of the states are swapped: $\el0 \leftrightarrow \eel0$ and $\el1 \leftrightarrow \eel1$. Then a two-tone, two-photon drive is used to move the two ions' errors into orthogonal superpositions of the COM and stretch modes while recovering the error-free logical state (Methods).
    Error bars are $1\sigma$ statistical uncertainties. Trials per point: 100 (raw) and 400 (detected) in \textbf{a}; 4,900 (raw and corrected) and 7,300 (detected) per offset, doubled at zero offset in \textbf{b}; 2,100 (raw and corrected) and 1,600 (detected) in \textbf{c}.}
    \label{fig:logical-gate}
\end{figure*}

\begin{figure*}
    \centering
    \includegraphics[width=1\linewidth]{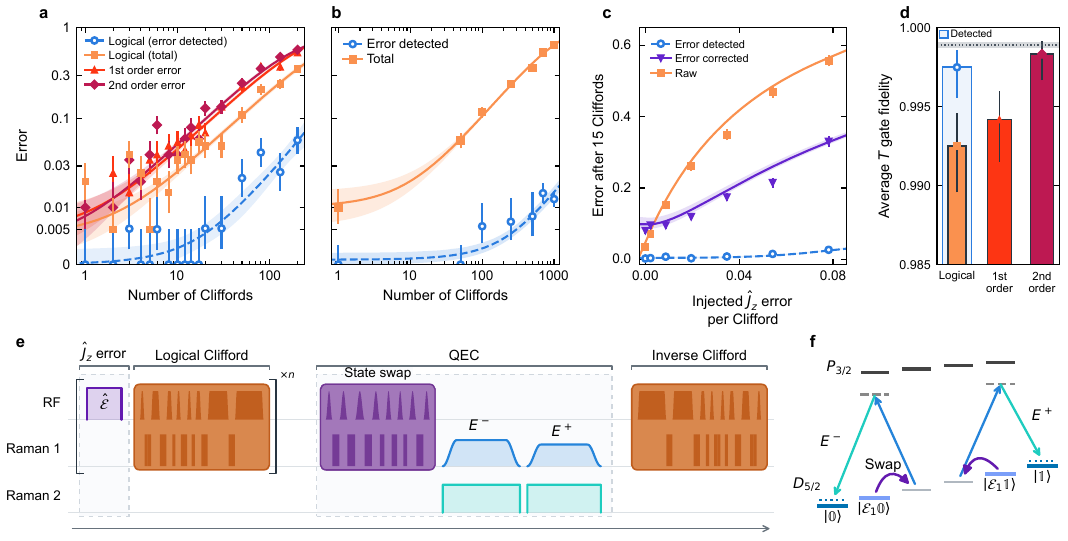}
    \caption{\textbf{Universal logical single-qubit gates and recovery from memory errors.}
    \textbf{a,} Randomized benchmarking of the error-tolerant Cliffords, with the state prepared and measured in the logical shell (total, and with error detection) and in the first and second error shells. \textbf{b,} Randomized benchmarking of the detection-only Cliffords, raw and with erasure checks during the sequence. 
    \textbf{c,} Logical error after 15 Cliffords interleaved with random $\jz$ rotations, versus the injected $\jz$ error per Clifford (Methods): raw, with recovery (corrected) and with error detection. 
    \textbf{d,} Average fidelity of the $\hat T_L$ gate in each shell over the six cardinal input states; for the logical shell the open bar shows the fidelity with error detection. The dotted gray line and band show the readout error and its $1\sigma$ uncertainty. 
    \textbf{e,} Benchmarking sequence: a $\jz$ error $\hat{\mathcal{E}}$ and a random logical Clifford are repeated $n$ times, followed by the optional recovery step and the inverting Clifford. 
    \textbf{f,} Level diagram of the recovery step. A state swap moves the first-error-shell population, and Raman transitions $E^\pm$ through $P_{3/2}$ return it to $\l0$ and $\l1$ while adding a motional quantum; dotted lines above $\l0$ and $\l1$ indicate the first excited motional state. 
    Curves are fitted benchmarking models (\textbf{a,b}) or calibrated predictions (\textbf{c}); bands are propagated $1\sigma$ fit uncertainties (\textbf{a}), central $1\sigma$ bootstrap intervals (\textbf{b}) and $1\sigma$ model uncertainties (\textbf{c}). Error bars are $1\sigma$ Wilson intervals (\textbf{a,b,d}) or counting standard errors (\textbf{c}). Trials per point: 200 in \textbf{a}; 400 at sequence lengths 1--500, 2,000 at 700 and 5,000 at 1,000 in \textbf{b}; 1,000 in \textbf{c}; 1,200 per shell in \textbf{d}, with 1,194 accepted in the logical shell.}
    \label{fig:clifford-rb}
\end{figure*}

\subsection{A multilevel entangling gate}
The geometric-phase gate familiar from qubits \cite{Lei03,Cen06,Lem13,Kam26} can be extended to the full spin manifold. A Hamiltonian of the form $(\hat{O}\otimes \hat{I}+\hat{I}\otimes \hat{O})$ applied in a space- and time-dependent manner to a two-ion crystal produces, after a closed loop in motional phase space, an entangling interaction $e^{-i\theta (\hat{O}\otimes \hat{I} \pm \hat{I} \otimes \hat{O})^2}$ up to an overall phase \cite{Roo07}, where $\hat{O}$ and $\hat{I}$ are single-ion internal-state operators, with $\hat{I}$ being the identity. We choose $\hat{O}=\jz$. Two Raman beams detuned by $\Delta$ from the $P_{3/2}$ manifold form a polarization gradient whose vector light shift exerts a force on ion $i$ proportional to its eigenvalue $m_i$ of $\jz$ (\cref{fig:logical-overview}c,d). With the ions spaced apart by a half-integer multiple of the effective wavelength, the two-ion component of the state with spin-projection difference $\Delta M=m_1-m_2$ then follows a phase-space loop scaled by $\Delta M$ and acquires a geometric phase $\Phi_{\Delta M}\propto \Delta M^2$ (\cref{fig:logical-overview}e; Supplementary Information). The resulting 36-dimensional operation is (up to overall phase) $e^{-i\theta (\jz\otimes \hat{I} - \hat{I} \otimes \jz)^2}$ and implements a logical $\widehat{XX}_L = e^{-i \hat{X}_L\otimes \hat{X}_L \pi/4}$ interaction at $\theta=\pi/2$ (up to local rotations and an overall phase). We apply it as two loops separated by a $\jy$ spin-echo pulse to refocus slowly varying errors odd in $\jz$, with adiabatic ramps of the drive amplitude and detuning (\cref{fig:logical-gate}d; Methods).

The choice $\hat{O}=\jz$ is what makes the gate compatible with the code. The spin-dependent force never changes the spin state $m_i$, so any imperfection in the motional part of the interaction, such as heating, a stray displacement, or a fluctuation in the force, acts on the internal states through a channel whose Kraus operators are polynomials in $\jz$ on each ion (after tracing out the motion). The only gate errors outside this set are processes that change the internal state, such as Raman scattering, and high-order terms with too large a power of $\jz$. The $\jz^2$ rotations produced by the tensor light shift of each beam also fall within the correctable set.

\subsection{Logical Bell states}
We prepare $\l0\!\l0$, apply the entangling gate, and remove a calibrated local $\jz^2$ rotation produced by the tensor light shifts, which leaves the ions in $(\l0\!\l0+i\l1\!\l1)/\sqrt{2}$ (Methods). We then decode into the $\jz$ basis and measure the populations and the parity \cite{Lei03,Bal14} (\cref{fig:logical-gate}a). In an optional error-detection step, population in the ground state (from $D_{5/2}$ decay or Raman scattering) and in the inner states $m\in\set{\pm3/2,\pm1/2}$ (from $\jz$ through $\jz^4$ errors) is flagged and the trial discarded. This process takes approximately $\SI{1}{\milli\second}$ and uses a multi-step \SI{729}{\nano\meter} deshelving sequence described in the Methods and in \cite{Bru26}.

The raw logical Bell-state fidelity is 92(1)\%. After error detection, the Bell-state fidelity rises to $99.8^{+0.1}_{-0.2}\%$ with 89.6(9)\% of trials accepted. Both values are close to the predictions of an error budget assembled from independent calibrations, namely 91(3)\% and 99.84(4)\%, respectively (Supplementary Information). The raw error is dominated by fluctuations in the pulse area of the two Raman beams, each of which produces a $\jz^2$ rotation on each ion, accounting for 5.0(3) percentage points of the predicted 9(3)\%, with a further 2.0 percentage points from unequal illumination of the two ions. This error is both correctable and detectable. With error detection, the residual error is dominated by state preparation and readout error estimated at 0.15(4)\%.

The logical Bell-state fidelity with no error correction or detection is lower than that of physical qubits in our system, measured at $98.3(1.1)\%$ with qubits encoded in the $\ket{\pm1/2}$ eigenstates of $\jz$ in the $D_{5/2}$ manifold. The higher raw error is due to the $\jz^2$ light shift errors that qubits do not have, as well as to the substantially larger phase-space excursions of the participating states in the logical gate, which increase sensitivity to a variety of motional errors \cite{Win97,Sut22b}. Only with error detection, which rejects about 10\% of trials, does the logical fidelity exceed the physical-qubit fidelity.

\subsection{Randomized benchmarking}
The error-detected Bell state fidelity is limited by state preparation and measurement (SPAM) errors, not entangling-gate errors. To quantify the fidelity of our entangling gates, we use two randomized-benchmarking techniques for entangling gates to extract SPAM-independent measures of the fidelity. The apparatus used in this work lacks individual addressing of the two ions, so both techniques apply only global pulses. Due to the approximately $6\%$ rejection per entangling gate, sequence lengths are limited to under 20 entangling gates, limiting the statistical precision. The randomized benchmarking measurements here were performed using different gate parameters than the Bell state fidelity measurements, with the differences described in the Methods.

Using subspace leakage error randomized benchmarking (SLERB) \cite{Sut25} we measure an error-detected infidelity per entangling gate of $0.84(10)\times 10^{-4}$ (\cref{fig:logical-overview}b) with $6.25(12)\%$ rejected per gate. This technique applies sequences of random two-qubit Cliffords on the two-level subspace $\set{\l0\!\l0,\l1\!\l1}$, built from entangling gates, single-qubit Cliffords, and single-qubit logical $\hat{Z}_L$ rotations (\cref{fig:methods-entangling-gate}g). Measuring in this way is sensitive not just to the entangling gates, but also to the single-qubit Cliffords, collective $\hat{Z}_L$ rotations, and error detection steps. We measure $0.6(1)\times 10^{-4}$ error per gate slot in sequences that omit the entangling gates, suggesting that the entangling gates themselves are only a small component of this error.

To extract an entangling gate error that is less sensitive to single-qubit rotations and error detection steps, we use symmetric stabilizer benchmarking (SSB) \cite{Tsa24}, measuring an error-detected infidelity per entangling gate of $0.19(13)\times 10^{-4}$ (\cref{fig:logical-overview}b; \cref{fig:methods-entangling-gate}f; Methods) with a rejection rate of $6.3(2)\%$ per gate. SSB uses a fixed number of random single-qubit Cliffords and erasure detection steps applied to both ions, and changes the number of entangling gates inserted into the gaps between the Cliffords (\cref{fig:methods-entangling-gate}h). The slope of the error as a function of the number of entangling gates thus suppresses the effect of single-qubit errors and measures the additional error introduced by the entangling gates, although the single-qubit error rejection is imperfect (Methods).

\subsection{Correcting entangling-gate errors}
To test whether local recovery (error correction) removes errors introduced during the gate, we implement an autonomous error correction step that returns population from the error shells to the logical shell and transfers the entropy of the error into two motional modes of the two-ion crystal \cite{Deb25,Li25d}. Global Raman sideband drives couple the two ions to orthogonal combinations of the COM and stretch modes, allowing a targeted error order to be locally removed from each ion (\cref{fig:logical-gate}e, \cref{fig:methods-qec-sequence}; Methods). Without individual addressing or sympathetic cooling, recovery itself adds error as predicted from independent calibrations (Methods). To demonstrate the benefit of recovery, we add controlled amounts of two types of noise during the gate and compare the Bell-state fidelity with and without recovery. 

In the first such experiment, we raise the power of one Raman beam above its calibrated value during the gate. This changes the phase-space trajectories' areas and results in uncompensated $\jz^2$ light shifts, both of which produce correctable $\jz^2$ errors (\cref{fig:logical-gate}b). At the largest power offset (a 1.7\% increase in power), applying recovery for $\jz^2$ errors raises the fidelity from 56(1)\% to 73(1)\%. With error detection the logical fidelity stays above 99\% over the full range, while the fraction of rejected trials increases from 11\% with no added noise to as much as 61\% at the highest errors. The crossover point at which error correction becomes beneficial is a 0.47(8)\% power offset.
In the second experiment we apply a spin-independent displacement with random phase to the motion midway through the first loop of the gate (\cref{fig:logical-gate}c). The kick leads to incorrect geometric phase for each $\Delta M$, which appears as a $\jz$-polynomial error on each ion. For a displacement $|\alpha|=1.11$, recovery raises the fidelity from 64(2)\% to 74(2)\% with a crossover point of a $|\alpha| = 0.7(1)$ kick. With error detection, the fidelity again stays above 99\%, with rejection rising from 9\% to 61\%. In both experiments the data agree with a model based on independent calibrations.

\subsection{Universal single-qubit gates}
Single-qubit operations can be constructed from combinations of RF rotations of the spin-$5/2$ manifold and a tensor light shift. The logical Pauli gates can be implemented as a single RF pulse resonant with the frequency difference between neighboring Zeeman states, which creates a $\jx$ or $\jy$ Hamiltonian. An RF $\pi$ pulse gives $e^{-i\pi\jy}=i\hat Y_L$ and $e^{-i\pi\jx}=i\hat Z_L$, and $e^{-i\pi\jz}=i\hat X_L$ is a virtual rotation implemented by advancing the phase of the RF oscillator by $\pi$ for subsequent pulses \cite{Mck16}. The logical $\sqrt{\hat{X}_L}$ gate uses the $\jz^2$ tensor light shift of a single Raman beam together with a virtual $e^{-i \jz \pi/2}$ rotation. This one-axis-twisting Hamiltonian over-squeezes the spin-coherent codewords into the cat states $(\l0\mp i\l1)/\sqrt{2}$, and it acts in the same way on the error shells \cite{Kit93,Li22i}. These gates do not generate the full Clifford group, so we construct the remaining Cliffords from numerically optimized sequences of RF rotations and $\jz^2$ pulses, which together are universal on the six-level manifold \cite{Oma21,Gio03,Pal03,Cha07b,Hei26} (Methods). A random logical Clifford requires on average 12.33 pulses and \SI{99.8}{\micro\second} (\cref{fig:methods-single-qubit}a--c).

Randomized benchmarking \cite{Mag10,Nie02,Che25e} gives an average error-tolerant logical Clifford fidelity of 99.78(2)\% on the logical shell and 99.61(2)\% averaged over the three shells, with the state prepared and measured in each shell in turn (\cref{fig:clifford-rb}a,e). We also construct a detection-only variant of the Clifford set in which errors are detectable but not necessarily correctable. With erasure checks inserted during the sequence, the undetected infidelity is $1.5(3)\times10^{-5}$ per gate with an erasure probability of $1.0 \times 10^{-3}$ per Clifford (\cref{fig:clifford-rb}b). These detection-only Cliffords only perform the desired operation on the outermost (logical) shell, which means they can be implemented in an average of just 3 pulses and \SI{50}{\micro\second}.

To complete the universal single-qubit gate set, we also need a non-Clifford gate such as the $T$ gate \cite{nielsen_quantum_2010}. We use similar optimal control techniques to construct pulse sequences that apply the error-tolerant $T$ gate as $\hat{T}_L = \hat{T}_0 \oplus\hat{T}_1\oplus\hat{T}_2$, where $\hat{T}_i$ implements a $T$ gate on shell $i$. We characterize the non-Clifford logical $T$ gate by measuring the overlap with the ideal output for each of the six cardinal states in each shell (\cref{fig:clifford-rb}d). The average fidelity is 99.5(1)\% across the three shells and 99.7(2)\% with error detection. These measurements are sensitive to SPAM errors (gray band). Benchmarking the $T$ gate more precisely would require more advanced techniques.

To test the error-tolerant Cliffords against memory errors, we interleave 15 random Cliffords with $\jz$ rotations applied through the RF antenna, with angles drawn from a normal distribution of width $\sigma$, and apply a single recovery step before the final inverting Clifford (\cref{fig:clifford-rb}c,e). A new random sequence is drawn for each trial. The recovery step maps the first error shell onto $m=\pm1/2$ while leaving the logical states at $m=\pm5/2$ and then drives motion-adding Raman sideband transitions $\Ket{\pm1/2}\to\Ket{\pm5/2}$ (\cref{fig:clifford-rb}f). Because the ion ideally starts in the motional ground state, the reverse transition is forbidden, so the step is one-way and deposits the entropy of the error in the motion. Similar to the entangling gate, single-qubit recovery costs more than it gains at the lowest noise levels, but above a crossover point it is beneficial. At an introduced error of $\sigma=\SI{0.125}{\radian}$ per Clifford, recovery increases the logical fidelity after 15 Cliffords from $74(1)\%$ to $88(1)\%$, and error detection gives a logical fidelity of $99.7(2)\%$. Nonzero motional occupation contributes $2.1(8)\%$ error of the total $3.9(8)\%$ recovery overhead at no injected noise, and fluctuating light shifts make up the bulk of the remaining error.

\section{Discussion}
\begin{figure*}
    \centering
    \includegraphics[width=1\linewidth]{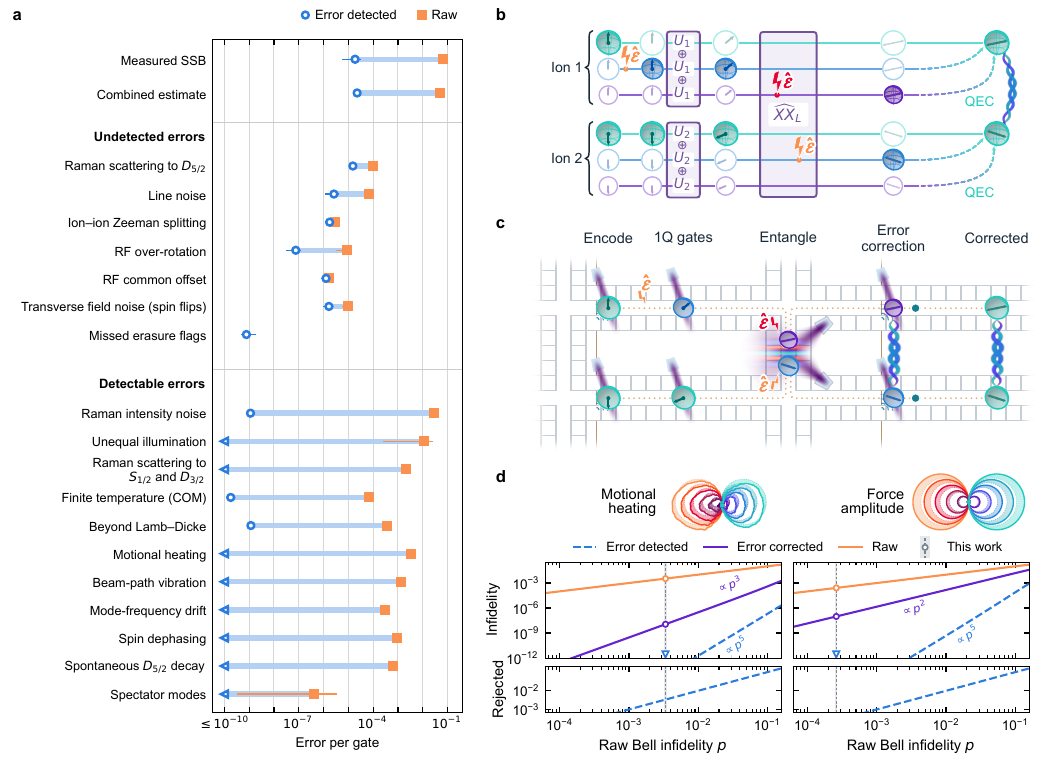}
    \caption{
    \textbf{Error budget and proposed architecture for spin-cat logical qubits.} 
    \textbf{a,} Error budget for logical entangling gates estimated from independent measurements of each term, grouped into undetectable and detectable errors. Markers denote estimates for raw infidelity contributions and error-detected infidelity contributions for each noise source. Triangles mark values $< 10^{-10}$; error bars show propagated uncertainty ranges. The measurement or simulation used for each term is described in the Methods or Supplementary Information.
    \textbf{b,} Circuit diagram for two $^{40}\mathrm{Ca}^{+}$ logical qubits across code ($\l0,\l1$) and syndrome ($\el0,\el1$, $\eel0,\eel1$) manifolds. Single-qubit rotations ($\hat{U}_1, \hat{U}_2$) precede a shared entangling gate $\widehat{XX}_L$ acting across all nine error shell combinations. Lightning bolts ($\hat{\mathcal{E}}$) mark dephasing errors; autonomous QEC returns populated syndrome rails to the code space. Where qubits are entangled, Schmidt vector arrows are shown in place of Bloch arrows.
    \textbf{c,} Five-zone logical QCCD processor corresponding to the circuit in \textbf{b}, including the logical gates, errors, and error correction. The logical QCCD architecture can be implemented in a standard QCCD surface trap with integrated light delivery.
    \textbf{d,} Top: phase-space trajectories showing exaggerated error during logical entangling gates due to motional heating (left) and incorrect force amplitude (right), with shaded areas representing missing or extraneous geometric phase. Middle: Bell infidelity versus raw infidelity $p$ for motional heating (left) and Raman force noise (right). Infidelities scale as $O(p^3)$ (heating) and $O(p^2)$ (force) under error correction, and $O(p^5)$ under detection. Bottom: corresponding rejected-shot fractions as a function of raw infidelity $p$. Vertical line and band: raw error of each mechanism from \textbf{a} ($1\sigma$).
    }
    \label{fig:limits}
\end{figure*}

By designing the entangling gate to only couple to the ions' motion through $\jz$, the same operator family that the spin-cat code protects against, nearly all entangling-gate errors can be corrected or detected, with $99.97(2)\%$ of all errors detected as erasures.
Imperfections of the ions' shared motion, which in conventional qubit gates lead to undetectable decoherence of their internal states, are instead mapped onto the shell structure of the code. They can then be corrected or flagged within each ion. Together with single-qubit gates that act equivalently on each shell, this gives an error-tolerant gate set in which the protection of a single-atom encoding applies to computation rather than just memory. This concept does not rely on details specific to trapped ions, and other multilevel systems coupled to a bosonic mode could inherit these protections as long as the coupling and the dominant error mechanisms can be made mutually diagonal. This potentially includes superconducting systems coupled to microwave cavities or atoms coupled to optical cavities.

The present error-correction demonstration has clear limitations that could be improved in future work. At the noise levels of our apparatus the recovery step adds more error than it removes, and local correction improves the logical fidelity only once additional noise is injected (\cref{fig:logical-gate}b,c and \cref{fig:clifford-rb}c). The cost of recovery is set mainly by the residual motional occupation and by the fidelity of the sideband transitions that carry the error entropy into the motion. Bringing this cost below the native gate error is required to reach break-even without added noise; this would require improved cooling to the motional ground state.
Recovery was also applied only once in each experiment, whereas sustained operation will require repeated rounds with the motion reset between them, for example by sympathetic cooling. Finally, the single-qubit gates are error tolerant only as complete operations, so errors that occur partway through them are not corrected by the spin-cat code and must be handled by an outer code.

Error detection is already in a stronger position. Excluding state preparation and readout, we measure undetected errors to be $0.19(13) \times 10^{-4}$ per gate, while $6.3(2)\%$ of trials are flagged, meaning that 99.97(2)\% of all gate errors are converted to erasures.
Residual logical errors at the $10^{-5}$ level, with nearly all errors appearing as erasures, are the regime in which erasure-tailored codes reach their highest thresholds \cite{Wu22,Kan22,Cha24b,baranes2026leveraging,Sah23b}. The price is the erasure rate itself, about 6.3\% per entangling gate, which is larger than many of those thresholds and must be reduced. It is dominated by light-shift errors (\cref{fig:limits}a), which can be reduced with better intensity stabilization and are absent for gates driven by magnetic-field gradients \cite{Osp08,Sut18,Oze06,Sri21}. Error correction before detection would also reduce the erasure rate. Apart from Raman scattering, the error-detected gate's error budget is limited not by entangling-gate-specific errors but by single-ion control and magnetic field errors. In field-insensitive qubits, single-qubit errors have been demonstrated at the $10^{-7}$ level \cite{Smi25}. This suggests that even lower logical error rates are possible with better control and stabilization.

These elements fit naturally into the quantum charge-coupled device (QCCD) architecture \cite{Win97,Pin20,Mos23}, as sketched in the circuit diagram and corresponding logical QCCD implementation in \cref{fig:limits}b,c. Each data ion carries a logical qubit that is moved, manipulated, entangled, and checked for erasures with the same lasers, RF fields, and motional modes used for a physical qubit. Active recovery requires a coolant ion to remove the entropy deposited in the motion, much as existing processors already pair each qubit ion with a cooling ion \cite{Mos23}, whereas detection alone requires no additional ions. The next steps in utilizing these codes are repeated rounds of recovery and detection with motional reset, and concatenation with an outer code designed for erasure-dominated noise. Combined with straightforward technical changes to reduce the erasure rate, single-ion encodings would then reduce the error rate seen by that outer code at no cost in ion number, which could substantially lower the overhead of fault-tolerant trapped-ion processors.

\section*{Methods}
\label{sec:methods}
\setcounter{figure}{0}
\setcounter{table}{0}
\renewcommand{\figurename}{Extended Data Fig.}
\renewcommand{\tablename}{Extended Data Table}
\renewcommand{\thefigure}{\arabic{figure}}
\renewcommand{\thetable}{\arabic{table}}
\renewcommand{\theHfigure}{methods.\arabic{figure}}
\renewcommand{\theHtable}{methods.\arabic{table}}
\crefalias{figure}{extendedfigure}

\subsection{Experimental setup}
\label{sec:methods-experiment}
These experiments are performed using a cryogenic surface-electrode ion trap that is described in \cite{Deb23,Deb25}. In addition to the previously described control using the narrow-linewidth \SI{729}{\nano\meter} laser that couples $D_{5/2}\leftrightarrow S_{1/2}$ and RF control driving transitions within $S_{1/2}$ ($\approx \SI{11.4}{\mega\hertz}$) and $D_{5/2}$ ($\approx \SI{6.8}{\mega\hertz}$), we introduce two free-space laser beams detuned several \si{\tera\hertz} from the $D_{5/2}\leftrightarrow P_{3/2}$ electric dipole transition, as shown in \cref{fig:logical-overview}. These beams are derived from a single titanium--sapphire laser and have $\approx\SI{40}{\milli\watt}$ each with a waist ($1/e^2$ diameter) of $\approx\SI{60}{\micro\meter}$.

\subsection{Spin-cat encoding and error tolerance}
\label{sec:methods-encoding}

\begin{figure*}
    \centering
    \includegraphics[width=1\linewidth]{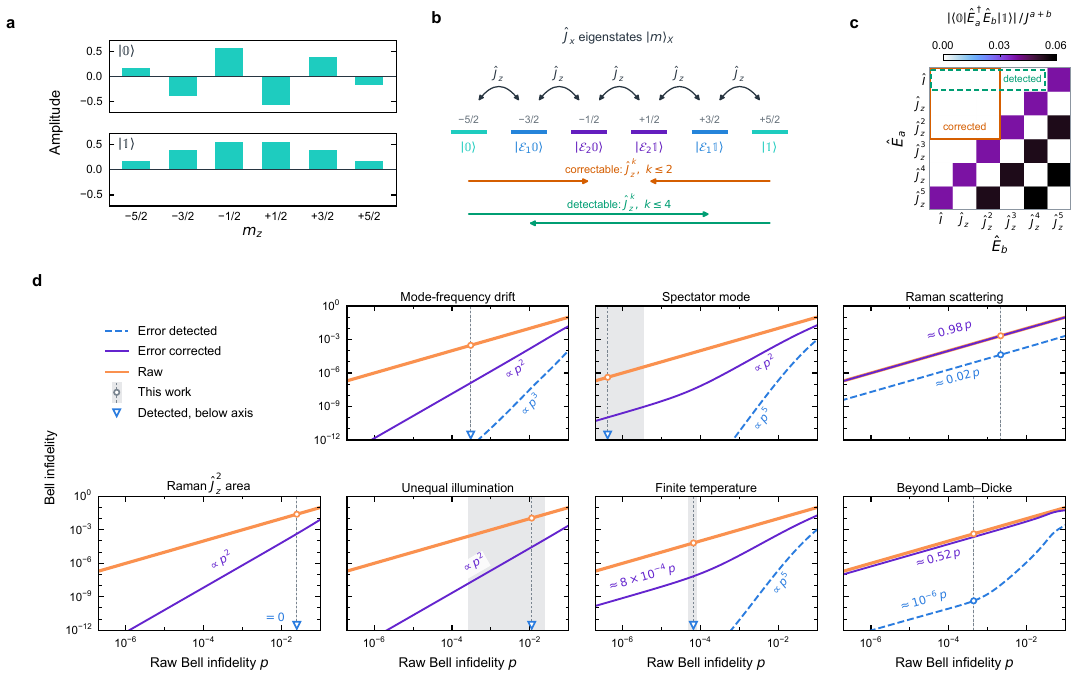}
    \caption{\textbf{Spin-cat codewords and their protection against entangling-gate errors.} 
    \textbf{a,} The amplitudes of the $\l0$ (top, \cref{eq:logical-codewords-0}) and $\l1$ (bottom, \cref{eq:logical-codewords-1}) codewords, which are the maximal eigenstates of $\jx$ and can be defined to be entirely real. 
    \textbf{b,} Protection against $\jz^k$ errors for $k\le 2$ (error correction) and $k\le 4$ (error detection). The $\jz^k$ errors move population between the error shells. 
    \textbf{c,} Knill--Laflamme condition matrix for $\jz^k$ errors on the spin-cat logical states. The matrix elements $\braket{\mathbb{0} \vert \hat {\mathcal{E}}_a^\dagger \hat {\mathcal{E}}_b \vert \mathbb{1}}$ for $\hat{\mathcal{E}}_{a,b} \in \set{\jz^k}$ vanish on the correctable block where $k\le2$ and on the top row where $\hat{\mathcal{E}}_a=\hat{I}$ and $\hat{\mathcal{E}}_b=\jz^k$ with $k\le4$. 
    \textbf{d,} Calculated logical Bell-state infidelity for entangling-gate error sources demonstrating the degree of error correction or error detection that is possible for each error type, assuming a spin-echo pulse is used during the gate.}
    \label{fig:methods-codewords}
\end{figure*}

No error-correcting code can correct or detect every possible error in a finite Hilbert space \cite{Kni97,Oma24b}, so a code should be matched to the errors that dominate in practice \cite{Gro20}. In trapped-ion qubits the dominant errors often arise during idling \cite{Rya21,Pae24} or entangling operations \cite{Gae16,Mos23,Eve23,Eve26}, both of which can manifest as dephasing. Fluctuations of the magnetic field, of the light shifts from the Raman beams, of the spin-dependent force, and of various properties of the shared motional mode all enter as polynomials in $\jz$. Spin-cat codes are designed for this error model \cite{Gro20,Oma24b,Deb25}. When memory or entangling-gate errors dominate over other error sources, a code designed for those errors can give a disproportionate reduction in the logical error rate \cite{Ali07c,Tuc18}. Other error mechanisms, such as control errors during logical single-qubit gates, can then be corrected by concatenating the spin-cat code with an outer code.

\textit{Code space and error subspaces.} The codewords are $\l0=\Ket{-5/2}_X$ and $\l1=\Ket{+5/2}_X$. In the $\jz$ basis, these are (\cref{fig:methods-codewords}a)
\begin{align}
    \begin{aligned}
        \l0 = \frac{1}{4\sqrt{2}}\Big(&
            \Ket{-5/2} - \sqrt{5}\Ket{-3/2}
            + \sqrt{10}\Ket{-1/2}\\
            &- \sqrt{10}\Ket{+1/2}
            + \sqrt{5}\Ket{+3/2} - \Ket{+5/2}
        \Big),
    \end{aligned}
    \label{eq:logical-codewords-0}\\[4pt]
    \begin{aligned}
        \l1 = \frac{1}{4\sqrt{2}}\Big(&
            \Ket{-5/2} + \sqrt{5}\Ket{-3/2}
            + \sqrt{10}\Ket{-1/2}\\
            &+ \sqrt{10}\Ket{+1/2}
            + \sqrt{5}\Ket{+3/2} + \Ket{+5/2}
        \Big).
    \end{aligned}
    \label{eq:logical-codewords-1}
\end{align}
The $k$-th error subspace is defined by orthogonalizing the states generated by $k$ successive applications of $\jz$ to the codewords, which gives $\el0=\Ket{-3/2}_X$ and $\el1=\Ket{+3/2}_X$ for $k=1$, and $\eel0=\Ket{-1/2}_X$ and $\eel1=\Ket{+1/2}_X$ for $k=2$. In the $\jx$ eigenbasis, $\jz$ couples only states whose eigenvalues differ by one, so $\braket{\mathbb{0}|\jz^k|\mathbb{1}}=0$ for $k<5$. A rotation by $\pi$ about the $z$ axis maps $\jx\to-\jx$ and leaves $\jz$ unchanged, which exchanges the two codewords and gives $\braket{\mathbb{0}|\jz^k|\mathbb{0}}=\braket{\mathbb{1}|\jz^k|\mathbb{1}}$ for all $k$. The Knill--Laflamme conditions are therefore satisfied for the error set $\{\hat{I},\jz,\jz^2\}$, whose pairwise products contain powers of $\jz$ up to four. Errors up to $\jz^2$ can thus be corrected, and errors up to $\jz^4$ can be detected (\cref{fig:methods-codewords}b,c).

\textit{Error tolerance.} We call an operation error tolerant if it does not convert a correctable ($\jz^k$, $k\le2$) error into an error that the code does not correct. Equivalently, a logical operator must act identically on the logical subspace and on each error subspace, without crosstalk between subspaces \cite{Kap17,Ma19b}. Writing each shell in the basis $\{\Ket{\mathcal{E}_i\mathbb{0}},\Ket{\mathcal{E}_i\mathbb{1}}\}$, a logical gate takes the form $\hat U_L=\hat U_0\oplus\hat U_1\oplus\hat U_2$ with identical matrices $\hat U_i$; a logical Hadamard gate, for example, is $\hat H_0\oplus\hat H_1\oplus\hat H_2$. We use two levels of tolerance. The entangling gate is diagonal in the $\jz$ basis and the spin-dependent force Hamiltonian commutes with $\jz$ on each ion, so errors that occur at any point during the entangling interaction remain correctable. The single-qubit gates are composed of several pulses (see the single-qubit gate section below); the complete sequence implements the same transformation on all three subspaces, but the intermediate pulses need not, so an error that occurs partway through a single-qubit gate is not in general correctable. Within this definition, the gate set is tolerant to first- and second-order dephasing errors for most of the operating time of a typical processor \cite{Mos23,Rans25} (dominated by transport and cooling times): between single-qubit gates and throughout entangling gates.

\textit{Geometric phase gate errors.} All high-fidelity entangling gates with trapped ions have been performed with a variant of the geometric phase gate (either in the $\hat{\sigma}_z$ basis \cite{Lei03} or as a $\hat{\sigma}_x$ M{\o}lmer--S{\o}rensen gate \cite{Sre00,Roo07}, where $\hat{\sigma}_k$ are the usual Pauli matrices). In both cases, spin-dependent forces push the $\hat{\sigma}_z$ or $\hat{\sigma}_x$ eigenstates through different phase space trajectories in the ions' shared motional phase space. Applied to an initial state $\sum_{m_1,m_2} c_{m_1,m_2} \Ket{m_1,m_2} \otimes \Ket{0}_\mathrm{mot}$, these spin-dependent displacements create states of the form
\begin{align*}
    \Ket{\Psi(t)} = \sum_{m_1,m_2} \Big(c_{m_1,m_2} & e^{i\Phi_{m_1,m_2}(t)} \Ket{m_1,m_2} \\
    &\otimes \hat{D}(\alpha_{m_1,m_2}(t))\Ket{0}_\mathrm{mot}\Big), 
\end{align*}
where $m_i$ is the internal state of ion $i$ in the $\hat{\sigma}_k$ basis, $\hat{D}(\alpha)$ is the displacement operator, and $\alpha_{m_1,m_2}(t)$ is the phase-space displacement of state $\Ket{m_1,m_2}$ at time $t$. Each spin state $\Ket{m_1,m_2}$ picks up a geometric phase
\begin{align*}
    \Phi_{m_1,m_2}(t) = \operatorname{Im} \left[ \int_0^t \alpha_{m_1,m_2}^*(t') \dot\alpha_{m_1,m_2}(t') \, dt'\right]
\end{align*}
proportional to the area of the trajectory of that state through phase space.
At the end of the gate ($t = t_{\mathrm{end}}$), any errors related to the application of the entangling gate that do not involve spin flips in the $\hat{\sigma}_k$ basis (i.e., excluding Raman scattering or spontaneous decay errors) manifest as either an incorrect value of $\Phi_{m_1,m_2}(t_{\mathrm{end}})$ or as residual spin-motion entanglement if $\alpha_{m_1,m_2}(t_{\mathrm{end}})$ is not equal for all $m_1,m_2$. Incorrect values of $\Phi_{m_1,m_2}(t_{\mathrm{end}})$ are diagonal in the $\hat{\sigma}_k$ basis and therefore act directly as phase errors. Lack of closure of $\alpha_{m_1,m_2}$ also manifests as a dephasing error when looking only at the internal state of the ions: tracing out the motion scales the off-diagonal spin coherence $\Ket{m_1,m_2}\Bra{m_1',m_2'}$ by the motional overlap
\begin{align*}
    \Braket{\alpha_{m_1',m_2'} | \alpha_{m_1,m_2}} = \exp\Big( &-\frac{1}{2}\left|\alpha_{m_1,m_2} - \alpha_{m_1',m_2'}\right|^2 \\
    &+ i\operatorname{Im}\left[\alpha_{m_1',m_2'}^*\alpha_{m_1,m_2}\right] \Big),
\end{align*}
which leaves the $\hat{\sigma}_k$-basis populations unchanged ($m_1=m_1', m_2=m_2'$) while damping the off-diagonal coherences. Thus, if a code that corrects dephasing errors is used with a proper generalization of the qubit geometric phase gates, motional errors can be corrected locally.

\subsection{Error detection}
\label{sec:methods-error-detection}
The spin-cat codewords can reject a wide range of errors using error detection. To do so, we detect any population that has either left the code space entirely (via decay or scattering to the $S_{1/2}$ or $D_{3/2}$ manifolds) or that is in either of the error shells spanned by $\el0,\el1,\eel0,\eel1$.

Detecting the error shell states requires decoding the codeword into the $\jz$ basis such that the logical amplitudes are on the $\Ket{\pm5/2}$ states. This removes the spin-cat code's protection from $\jz$ errors and makes the logical state susceptible to dephasing errors. Additionally, the pulses used to measure the error shell populations can introduce light shifts that can also cause logical dephasing errors. To minimize these effects in the randomized benchmarking experiments, we minimize the amount of time that the information is in the unprotected basis and apply a spin-echo RF pulse halfway through the detection sequence to refocus logical dephasing errors. The deshelving sequence (including spin-echo pulse) takes approximately \SI{400}{\micro\second}. The fluorescence detection step of error detection is performed after re-encoding into the protected basis and takes approximately \SI{560}{\micro\second}.

For all deshelving steps, we employ a multi-pulse sequence where spectrally resolved \SI{729}{\nano\meter} quadrupole transition pulses drive population from a selected state $\ket{m}$ to the $S_{1/2}$ manifold. The \SI{729}{\nano\meter} pulses alternate with \SI{397}{\nano\meter} pulses coupling the $S_{1/2}$ to $P_{1/2}$, which results in optical pumping of the population to the metastable $D_{3/2}$ manifold \cite{Bru26}. This enables multiple \SI{729}{\nano\meter} deshelving pulses for each $m$ state without re-excitation to $D_{5/2}$, removing sensitivity to the fidelity of the deshelving pulses. Then, fluorescence detection with the \SI{397}{\nano\meter} and \SI{866}{\nano\meter} lasers is used to detect any deshelved (erasure) population that is in either the $S_{1/2}$ or $D_{3/2}$ manifolds.

In the logical Bell state experiments, the fluorescence detection is performed immediately after the deshelving sequence. In the randomized benchmarking experiments, however, the fluorescence detection is delayed until the end of the sequence, taking advantage of the ability to pump population into the $D_{3/2}$ manifolds. In those cases, any population in $D_{3/2}$ is pumped back into the $S_{1/2}$ manifolds before the next logical gate is applied to prevent Raman scattering from the $D_{3/2}$ manifold back into the $D_{5/2}$ manifold.

\subsection{Entangling gate}
\label{sec:methods-gate}

\begin{figure*}
    \centering
    \includegraphics[width=1\linewidth]{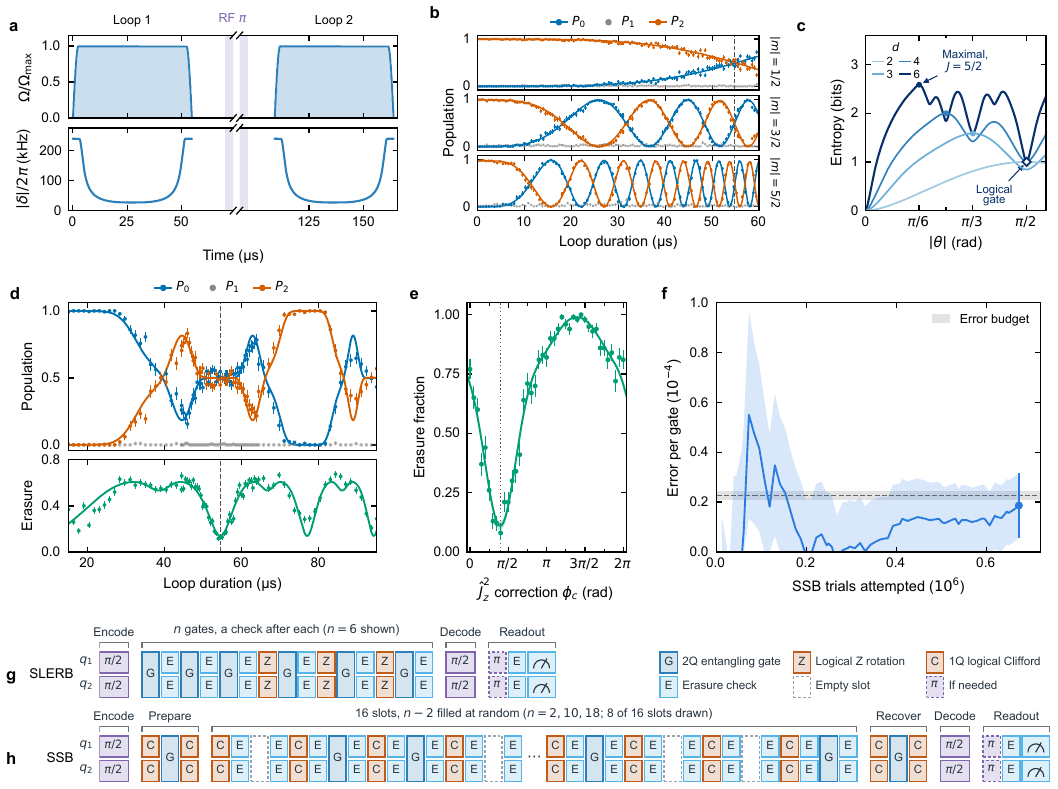}
    \caption{\textbf{Implementation and calibration of the multilevel entangling gate.} 
    \textbf{a,} Pulse shape used to implement the entangling gate, based on \cite{Sut23b,Bow25,Hug25}. 
    \textbf{b,} Measurements of entangling interactions of two-level systems where each ion is prepared in an equal superposition of the $\Ket{\pm m}$ eigenstates of $\jz$, with $\Delta M=-2m,0,+2m$ collective spin states. The population oscillations can be used to measure the gate time of the logical qubit gate, which occurs when $|m|=1/2$ states have a geometric phase of $\pi/2$, $|m|=3/2$ states have a geometric phase of $9\pi/2$, and $|m|=5/2$ states have a geometric phase of $25\pi/2$, denoted by the dashed vertical line. 
    \textbf{c,} This entangling interaction can be used both for logical qubit gates and for maximally entangling qudit gates. The plot shows the entanglement entropy created by an entangling gate with interaction angle $\theta$ on a $d$-dimensional system, where the logical qubit gate is at $\lvert\theta\rvert=\pi/2$ and the maximally entangling gate for $d=6$ is at $\lvert\theta\rvert=\pi/6$. In this plot, the ion is prepared in an equal-superposition state.
    \textbf{d,} Population scan of the entangling gate as a function of loop duration with erasure conversion. The gate time occurs where the populations are equal and the erasure fraction is minimized, shown by the vertical dashed line. 
    \textbf{e,} At the optimal gate time, the final $\jz^2$ correction pulse time is scanned to minimize the erasure fraction that results from uncompensated light shifts.
    \textbf{f,} SSB measurement campaign, showing the measured error per gate and $1\sigma$ uncertainty as a function of the cumulative number of attempted trials, plus the final measurement value and $1\sigma$ uncertainty.
    \textbf{g,} Subspace leakage error randomized benchmarking (SLERB) pulse sequence. The logical pair is encoded in $\l0\!\l0$, followed by $\ell$ random Cliffords ($\ell=3$ shown) comprised of $n$ entangling gates, each followed by an erasure check. The number of entangling gates $n$, the number of single-qubit gates, and the number of detections grows with $\ell$.
    \textbf{h,} Symmetric stabilizer benchmarking pulse sequence. The ions are encoded and prepared in a symmetric stabilizer state followed by a sequence with a fixed number of random Cliffords and erasure checks, with variable numbers of entangling gates inserted randomly into the slots between Cliffords. Only 8 of the 16 slots are shown.
    }
    \label{fig:methods-entangling-gate}
\end{figure*}

Our entangling gate relies on a Hamiltonian that applies a force linear in the difference of the quantum numbers $\Delta M=m_1-m_2$. This interaction has favorable properties in terms of error correction and detection and is straightforward to apply with two laser beams or with magnetic field gradients, as the magnetic field coupling to a linear Zeeman manifold is strictly $\mathrm{SU}(2)$-symmetric, enabling the application of $\jz \otimes \jz$ or $\jy \otimes \jy$.

\textit{Beam geometry and Hamiltonian.} Two Raman beams with wavevectors $\vec{k}_1=k(-\boldsymbol{e}_x+\boldsymbol{e}_y)/\sqrt{2}$ and $\vec{k}_2=k(\boldsymbol{e}_x+\boldsymbol{e}_y)/\sqrt{2}$ illuminate the ions (\cref{fig:logical-overview}c); $\boldsymbol{e}_x$ is along the trap axis and $\boldsymbol{e}_y$ is the in-plane axis perpendicular to it, with the quantization axis along $\boldsymbol{e}_z$ normal to the trap surface. Both beams are detuned by $\Delta$ from the $D_{5/2}\leftrightarrow P_{3/2}$ transition at \SI{854}{\nano\meter}, where $\Delta \approx \SI{5.15}{\tera\hertz}$ for the randomized benchmarking experiments and $\Delta\approx\SI{1.776}{\tera\hertz}$ in the Bell state fidelity measurements. The beams are linearly polarized in the plane of the trap, with $\boldsymbol{\epsilon}_1=(\boldsymbol{e}_x+\boldsymbol{e}_y)/\sqrt{2}$ and $\boldsymbol{\epsilon}_2=(\boldsymbol{e}_x-\boldsymbol{e}_y)/\sqrt{2}$. The frequency difference $\omega$ of the beams is set near the frequency $\omega_\mathrm{COM}$ of the axial center-of-mass mode. After the common optical frequency is factored out, the positive-frequency field at the ion is
\begin{align*}
    \vec{E}^{(+)}(x,t)\propto\sin(k_x x-\omega t/2)\,\mathbf{e}_{-} - \cos(k_x x-\omega t/2)\,\mathbf{e}_{+},
\end{align*}
with $k_x=k/\sqrt{2}$. The $\sigma^\pm$ components produce a spatially varying vector light shift \cite{steck2026quantum}
\begin{align*}
    \Delta E^{(1)} &= \alpha_{\mathrm{v}}\jz\left(\lvert\vec{E}^{(+)}\cdot(\mathbf{e}_{+})^*\rvert^2-\lvert\vec{E}^{(+)}\cdot(\mathbf{e}_{-})^*\rvert^2\right)\nonumber\\
    &\propto \alpha_{\mathrm{v}}\jz\cos(\Delta k\,x-\omega t),
\end{align*}
where $\alpha_{\mathrm{v}}$ is the vector polarizability at this detuning and $\Delta k=2k_x=\sqrt{2}\,k$ is the magnitude of the effective wavevector, which lies along the trap axis. The spherical basis vectors are $\mathbf{e}_{\pm} = \mp (\mathbf{e}_x \pm i \mathbf{e}_y)/\sqrt{2}$. The two-ion Hamiltonian is therefore
\begin{align}
    \hat{H}=\Omega(t)\left(\jz\otimes \hat{I}-\hat{I}\otimes\jz\right)\cos(\Delta k\,\hat{x}-\omega t), \label{eq:ham-eff}
\end{align}
with time-dependent Rabi frequency $\Omega(t)$. The scalar and tensor light shifts are spatially uniform in this geometry, so they do not contribute to the force; the tensor part produces the local $\jz^2$ rotation that is compensated as described below. In the interaction picture with respect to the motion,
\begin{align*}
    \hat{H}_I=&\frac{\Omega}{2}\left(\jz\otimes \hat{I} - \hat{I}\otimes\jz\right)\hat{D}\!\left(i\eta_{\mathrm{COM}} e^{i \omega_\mathrm{COM} t}\right)e^{i(\phi_{\mathrm{R}}-\omega t)}\\&\qquad+\text{h.c.},
\end{align*}
where $\hat{D}$ is the displacement operator, and $\phi_{\mathrm{R}}$ is the relative phase of the two beams. For the two-ion COM mode, $\eta_{\mathrm{COM}} = \Delta k \sqrt{\hbar / (4 M_\mathrm{ion} \omega_\mathrm{COM})} \approx 0.061$ is the Lamb--Dicke parameter, where $M_\mathrm{ion}$ is the mass of one ion. The single-ion Lamb--Dicke parameter in the same trap is 0.086. We also define the detuning of the applied frequency $\omega$ from the motional mode frequency $\omega_\mathrm{COM}$ as $\delta=\omega - \omega_\mathrm{COM}$. The component of the two-ion state with two-ion projection $\Delta M=m_1-m_2$, which ranges from $-5$ to $+5$, follows a phase-space trajectory scaled by $\Delta M$, so that its mean phonon number and geometric phase are larger by $\Delta M^2$ than in a qubit gate. In these experiments the COM mode frequency is $\omega_\mathrm{COM}\approx2\pi\times \SI{1.80}{\mega\hertz}$. The full-power two-beam Rabi frequency is $\Omega \approx 2\pi \times \SI{298}{\kilo\hertz}$. 

\textit{Pulse shaping.} Because the phase-space excursions are up to five times larger than in a qubit gate, the gate is more sensitive to fluctuations of the trap frequency and the magnetic field, to effects beyond the Lamb--Dicke regime, and to laser phase noise, although most of the resulting errors are detectable and correctable. To minimize these errors, we apply two adiabatic gate loops based on the gates described in \cite{Hug25,Bow25,Sut23b}, each with a loop entangling phase of $\theta_\mathrm{loop}=\pi/4$, separated by a $\jy$ spin-echo pulse (\cref{fig:methods-entangling-gate}a). The intensity of one laser is switched on and off with $\sin^2$ ramps that occupy a fraction $r \approx 0.02$ of each loop of duration $t_\mathrm{loop}=\,$\SI{118}{\micro\second} (for the SLERB and SSB measurements; it was about \SI{47}{\micro\second} in the Bell-state fidelity measurements), and the detuning is ramped adiabatically during the gate according to the profile used in \cite{Hug25} with a maximum detuning of $\approx 2\pi \times 3/t_\mathrm{loop}$ and a minimum detuning of $\approx 2\pi\times 1/t_\mathrm{loop}$. This reduces the sensitivity to frequency fluctuations of the trap and of the lasers, and suppresses higher-order terms of the interaction that would otherwise reduce the fidelity substantially. A detailed pulse sequence with delay times specified for the Bell state fidelity measurements is included in the Supplementary Information.

\textit{Calibration.} The entangling gate can be calibrated by first performing two-level system entangling gates as shown in \cref{fig:methods-entangling-gate}b, where the logical gate occurs when all two-level systems are entangled. Preparing the logical state instead leads to a population scan as shown in \cref{fig:methods-entangling-gate}d, where the gate time minimizes the erasure fraction and leads to equal $\l0\!\l0$ and $\l1\!\l1$ population. This relies on properly calibrating the $\jz^2$ correction angle (\cref{fig:methods-entangling-gate}e), and an iterative process may be used to calibrate both the gate time and correction angle.

\textit{Error structure.} The fidelity of the entangling operation is defined on the internal states of the two ions, with the motional state traced out. The spin-dependent forces used in geometric-phase gates do not change the spin state, so their action on the internal states can be described as an error channel whose Kraus operators are expandable in powers of the local $\jz$ operator on either ion. Any motional error in which the power of $\jz$ on each ion is at most $J-1/2$ is therefore corrected by the spin-cat code, and any error with power up to $2J-1$ is detectable (\cref{fig:methods-codewords}d). The only gate errors that the code does not correct are errors that change the internal state of the ion and high-order terms with too large a power of $\jz$. The same reasoning applies to a M{\o}lmer--S{\o}rensen-type interaction: a $\jz$-basis spin-cat code allows correction or detection of all low-order motional errors of a $\jy\otimes\jy$ gate.

\textit{State preparation and light-shift compensation.} We prepare $\l0\!\l0$ by initializing each ion in the $-5/2$ eigenstate of $\jz$ and applying an RF $e^{-i \jy \pi/2}$ rotation. The two gate loops surrounding the spin-echo pulse then produce the state $e^{-i\phi_{\mathrm{LS}}\jz^2}(\l0\!\l0+i\l1\!\l1)/\sqrt{2}$, where $e^{-i\phi_{\mathrm{LS}}\jz^2}$ denotes a $\jz^2$ rotation of angle $\phi_{\mathrm{LS}}$ applied to each ion. This rotation arises from the single-beam tensor light shifts. We calibrate it and remove it with a $\jz^2$ rotation of angle $\phi_c$ chosen such that $\phi_{\mathrm{LS}}+\phi_c=0\pmod{2\pi}$, which leaves the ions in $(\l0\!\l0+i\l1\!\l1)/\sqrt{2}$ (\cref{fig:methods-entangling-gate}d,e). The state is then decoded into the $\jz$ eigenbasis, and the fidelity is obtained from the joint populations and coherences \cite{Bal14,Lei03}.

\textit{Partial state tomography.} The entangling gate ideally results in the logical entangled state $(\l0\!\l0 + i \l1\!\l1)/\sqrt2$. The density matrix for this state has four real parameters: the two-ion population of the $\l0$ state $P_{00}$, the two-ion population in the $\l1$ state $P_{11}$, and the magnitude $C/2$ and phase of the coherences between the logical states. Measuring those parameters is sufficient to determine the phase-corrected fidelity to the entangled state as $(P_{00}+P_{11}+C)/2$. The populations are measured simply by checking both the $\l0$ and $\l1$ states (by decoding with RF $\pi/2$ pulses and then deshelving with \SI{729}{\nano\meter} pulses). To measure the coherences, we similarly decode into the $\jz$ basis with a RF $\pi/2$ pulse, followed by the application of a $\pi/2$ pulse with variable phase that acts between the $\ket{+5/2}$ and $\ket{-5/2}$ states on both ions, and then deshelve and measure the $\ket{-5/2}$ state. By calculating the parity $\braket{\hat{\Pi}} = P_{0} + P_{2} - P_{1}$, where $P_{i}$ is the probability of measuring $i$ ions bright, it is possible to extract the magnitude and phase of the coherences from the amplitude of the oscillation of the parity (\cref{fig:logical-gate}a) \cite{Lei03}. We combine 1,200 trials of the populations with 400 trials per point of the coherences to measure an error-detected logical Bell state fidelity of $99.8^{+0.1}_{-0.2}\%$, and 1,100 population measurements and 100 parity trials per point to measure $92(1)\%$ Bell state fidelity without error detection.

\textit{Error detection.} The detection step is performed after decoding and before the final state readout. It flags population in the ground state, which results from $D_{5/2}$ decay or Raman scattering, and in the interior states $m\in\{-3/2,-1/2,+1/2,+3/2\}$, which results from $\jz$ through $\jz^4$ errors. The detection sequence is described in \cref{sec:methods-error-detection}.

\subsection{Local recovery}
\label{sec:methods-recovery}

\begin{figure*}
    \centering
    \includegraphics[width=1\linewidth]{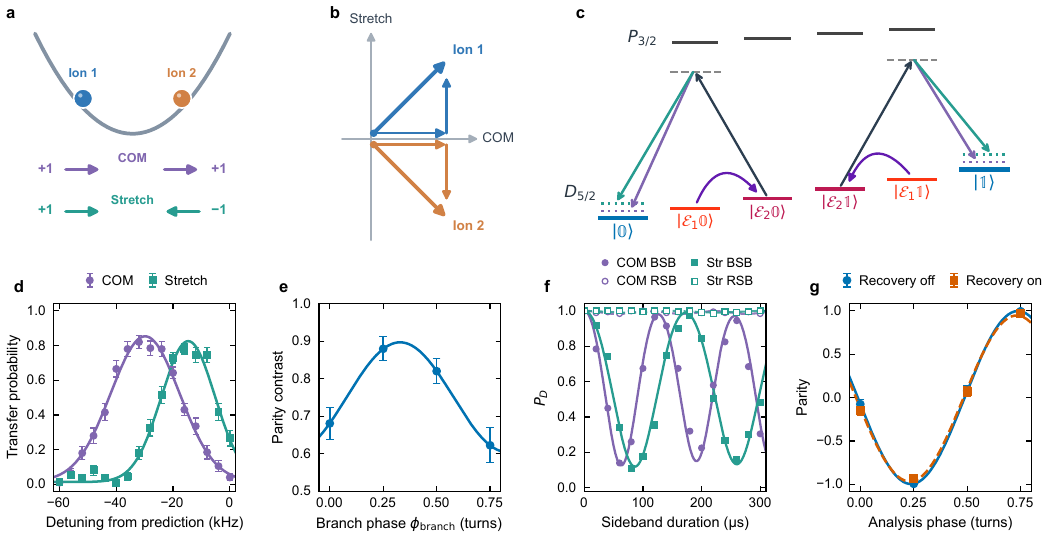}
    \caption{\textbf{Local error correction with motional modes.} \textbf{a,} Two ions in a shared potential have center-of-mass (COM) and stretch motional modes with mode participation factors listed in the figure. \textbf{b,} By driving the COM and stretch modes simultaneously, two orthogonal combinations of COM and stretch can be created, each of which interacts with only one of the ions, allowing two-ion error correction with only global control. \textbf{c,} Energy-level diagram showing the correction of first-order errors. These are first swapped to the $\eel0$ and $\eel1$ states. Then a bichromatic drive is applied, transferring any errors into the COM and stretch superpositions shown in \textbf{b}. \textbf{d,} Spectroscopy of the COM and stretch Raman transitions used for two-ion local error correction. \textbf{e,} Scan of the Bell-state parity as a function of the phase between the COM and stretch branches, which is used to ensure orthogonality of the motional states for ion 1 and 2. 
    \textbf{f,} Thermometry of the COM and stretch modes after sideband cooling, which places a limit on the maximum recovery fidelity. 
    \textbf{g,} Logical Bell-state parity scan with recovery on and off at no injected noise showing the reduction in parity due to the recovery sequence.}
    \label{fig:methods-qec-sequence}
\end{figure*}

\begin{figure*}
    \centering
    \includegraphics[width=1\linewidth]{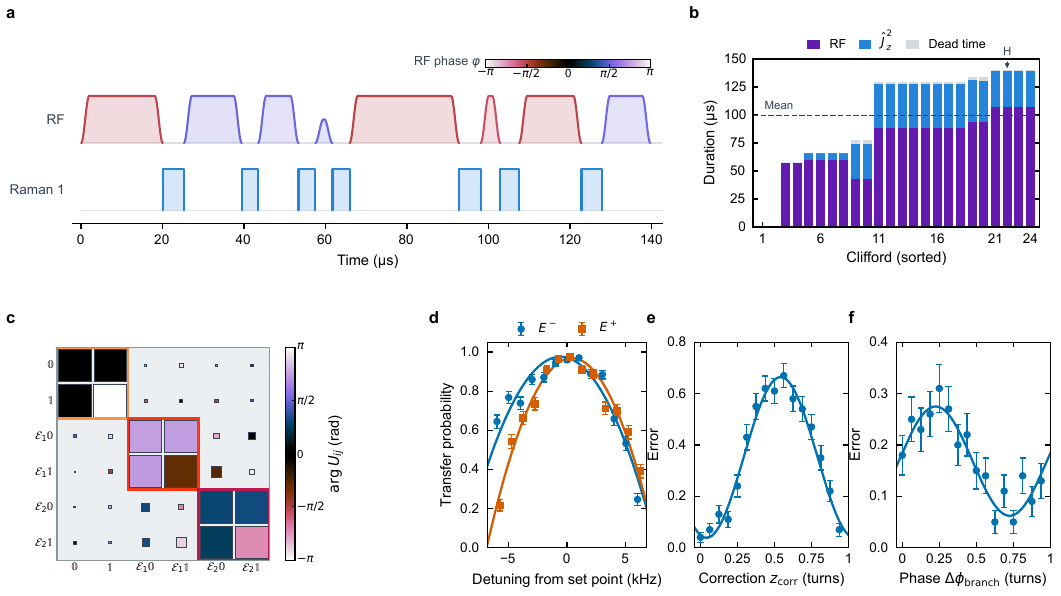}
    \caption{\textbf{Construction of logical single-qubit Cliffords and calibration of single-ion recovery.} \textbf{a,} Example Clifford optimal-control sequence performing a logical Hadamard gate. \textbf{b,} Distribution of pulse durations for performing each of the 24 Clifford gates, showing time spent on RF pulses, $\jz^2$ light shifts, and dead time between pulses. 
    \textbf{c,} Numerical reconstruction of the 6-dimensional unitary for the Hadamard gate. Relative phases are present between logical subspaces, but within each subspace equivalent Hadamards are applied up to an overall phase. The size of each square is proportional to the magnitude of the matrix element, and the color represents the phase of the matrix element. 
    \textbf{d,} Measurement of the two-photon sideband transition frequencies for performing the motional correction (see \cref{fig:clifford-rb}f). 
    \textbf{e,} Calibration of a virtual phase to correct light shifts on the logical states introduced by the two-photon correction sequence. 
    \textbf{f,} Measurement of the relative phase between the $E^+$ and $E^-$ arms of the motional correction on a state that is initially prepared entirely in the $\mathcal{E}_1$ subspace. }
    \label{fig:methods-single-qubit}
\end{figure*}

Recovery is autonomous in that it is measurement-free \cite{Sar05,Rei17} and the same operations are applied to all ions without knowledge of whether or not an error has occurred on any particular ion. It maps the error shells onto $\jz$ eigenstates that can be returned to the logical states by motion-adding Raman sideband transitions, and the entropy of the error can be carried away by cooling of the motional mode rather than by a measurement \cite{Deb25,Li25d}. For the single-qubit experiments, a state remapping operation places the population of $\el0$ and $\el1$ in the $\pm1/2$ eigenstates of $\jz$ and leaves the $\l0,\l1$ superposition in the $\pm5/2$ eigenstates. The two Raman beams then drive two-photon $\Ket{\pm1/2}\mapsto\Ket{\pm5/2}$ transitions that add one motional quantum (\cref{fig:clifford-rb}f, \cref{fig:methods-single-qubit}d--f); the reverse, motion-subtracting transition is ideally forbidden because the ion starts in the motional ground state. The fidelity of recovery is limited by the achievable motional occupation, though errors introduced through a nonzero thermal occupation are themselves detectable. Repeated recovery requires resetting the motion between rounds, for example by sympathetic cooling.

For the two-ion Bell-state experiments, the entropy is transferred into two motional modes of the crystal. This apparatus lacks the ability to address the two ions individually, which complicates the use of the motional modes for error recovery, as the COM and stretch modes are shared by the two ions with equal participation magnitudes (\cref{fig:methods-qec-sequence}a). 
If ion 1 had an error and ion 2 did not prior to recovery, using the COM mode with global pulses could at best result in a situation where ion 1 has a 25\% chance of error, ion 2 has a 25\% chance of error, and the COM mode has 50\% excitation. This is because driving a global sideband creates a three-state ladder connecting ion 1's error state to ion 2's error state. To avoid this, we drive both the COM and stretch sideband simultaneously. The two ions have the same participation in the COM mode, but the signs of their participation in the stretch mode are opposite. This means that there is a rotated motional basis where ion 1 and ion 2 couple independently to orthogonal combinations of modes (\cref{fig:methods-qec-sequence}b), accessed by applying the COM and stretch sidebands simultaneously with the appropriate phase relationship applied to the laser drives (\cref{fig:methods-qec-sequence}d,e).

For the single-ion recovery experiments, the recovery overhead at no noise is $3.9(8)\%$, of which $2.1(8)\%$ is from nonzero motional excitation of the axial mode used for the recovery and the remainder is likely from fluctuating light shifts. For the two-ion recovery, the reduction in fidelity at no added noise is $2.3(8)$ ($\jz^2$) and $7(2)$ ($\jz$, including swap) percentage points (\cref{fig:methods-qec-sequence}g). 
Calibrations give a survival probability of error-free states of $\approx96\%$ and a probability of correcting an error of $\approx60\%$.
This is primarily due to motional excitation of the COM and stretch modes during these experiments, with their excitations, $\bar{n}_\mathrm{COM}=0.036(11)$ and $\bar{n}_\mathrm{stretch}=0.020(8)$ (\cref{fig:methods-qec-sequence}f). The two-ion recovery sequence employed additional spin-echo pulses that reduced the error from fluctuating light shifts by subjecting both logical states to the same light shifts.

Each experiment corrects only a single error order, either $\jz$ or $\jz^2$. With no noise, the infidelity is dominated by $\jz^2$ errors. The $\jz^2$ errors naturally result in states that the two-photon transition can couple to the logical states where $m$ changes by 2 (\cref{fig:logical-gate}e, \cref{fig:methods-qec-sequence}c). However, $\jz$ errors differ by one unit of angular momentum from the logical states, so we first apply an internal state operation that maps $\ket{\pm 3/2}\mapsto\ket{\pm1/2}$, using the same optimal control techniques as the Cliffords (\cref{fig:clifford-rb}f, \cref{fig:methods-qec-sequence}c). Then the same $|\Delta m|=2$ Raman sidebands can be used to perform the correction of $\jz$ errors.

Additional details about the two-photon error recovery pulses are included in the Supplementary Information.

\subsection{Randomized benchmarking}
\label{sec:methods-rb}
The partial state tomography measurement is limited by state preparation and measurement error, so most of the $0.2\%$ infidelity is unrelated to the entangling gate itself. To properly benchmark the entangling gate, we can use randomized benchmarking, repeatedly applying the gate interleaved with random single-qubit operations and fitting the slope to determine the error per gate. There are two main limitations in the present case. First, the experimental apparatus does not have the capability to individually address the two ions, so all operations are global. SLERB \cite{Sut25} avoids this by benchmarking only the symmetric $\set{\l0\!\l0,\l1\!\l1}$ subspace. Second, most benchmarking techniques measure not only the entangling gate contribution to the error, but also to the single-qubit operations and (in our case) the erasure detection pulses. A measurement of the entangling-gate error that is less sensitive to the single-qubit error contributions can be obtained using SSB \cite{Tsa24}.

In the SLERB and SSB randomized benchmarking experiments, we use a Raman detuning $\Delta \approx \SI{5.15}{\tera\hertz}$ red of the \SI{854}{\nano\meter} transition in $^{40}\text{Ca}^+$ to decrease Raman scattering errors. At this detuning and power, the gate time is $t_\mathrm{loop}\approx\SI{118}{\micro\second}$ per loop, compared to $t_\mathrm{loop}\approx\SI{47}{\micro\second}$ per loop in the Bell state measurements. We use a detuning ramp from $2\pi \times 3/t_\mathrm{loop}$ to $2\pi \times 1/t_\mathrm{loop}$ and an amplitude ramp fraction $r=0.02$. The two loops are again split by an RF $\pi$ pulse of $\approx\SI{55}{\micro\second}$, giving a total time per added entangling gate of \SI{358.8}{\micro\second}.

\subsubsection{Subspace leakage error randomized benchmarking}
Subspace leakage error randomized benchmarking (SLERB) \cite{Sut25} measures the error from applying random Cliffords on the $\set{\l0\!\l0,\l1\!\l1}$ subspace as a function of the number of Cliffords. We apply $\ell=0,\ldots,5$ random Cliffords to this effective two-level subspace, followed by a final inversion Clifford. The compilation uses an average of $13/6$ entangling gates per Clifford. Among trials that are not flagged as erasures, there are three possible outcomes. The result is either the expected two-ion state, leakage out of the subspace into an odd-parity state, or a two-ion error where both ions are in the wrong subspace state. These results are jointly fitted via multinomial maximum likelihood. The data in \cref{fig:logical-overview}b are plotted as a function of the average number of entangling gates $n=(13/6)\ell$. The fit to the data gives an error $0.84(10)\times 10^{-4}$ and an erasure fraction of $6.3\%$ per gate. This comes from just over one million trials, of which approximately 627,000 were accepted, taken over about \SI{33}{\hour}. The runtime includes occasional calibrations of the gate time, the $\jz^2$ compensation pulse duration, RF frequency and pulse durations, \SI{729}{\nano\meter} frequencies and pulse durations, and feedforward correction strengths for the \SI{60}{\hertz} line noise \cite{Tat26}.

The SLERB error includes errors from the single-qubit rotations as well as from the error detection steps. During the SLERB measurement run, we performed comparison sequences with the same single-qubit rotations and erasure checks but omitted the entangling gates. This measured a comparable error per gate-free slot of $0.57(11) \times 10^{-4}$, which suggests that much of the measured SLERB error comes from the single-qubit rotations and error detection steps. This is consistent with the SSB measurement described below.

\subsubsection{Symmetric stabilizer benchmarking}
An entangling-gate fidelity that contains a smaller contribution from the errors introduced by single-qubit operations can be measured via symmetric stabilizer benchmarking \cite{Tsa24}, which uses a fixed number of single-qubit operations but a variable number of entangling gates. We perform an SSB sequence with between 2 and 18 entangling gates (\cref{fig:methods-entangling-gate}h), with the maximum number set by the decreasing acceptance fraction as the number of gates increases. The ions are ground-state cooled and prepared in one of twelve symmetric stabilizer states using single-qubit Cliffords and one entangling gate, followed by 16 rounds of gates, and then read out with single-qubit Cliffords and an additional entangling gate. Each round consists of a random single-qubit Clifford, an erasure check, a slot that may or may not be filled with an entangling gate, and a second erasure check. The erasure checks use a sequence of pulses that results in the error states being optically pumped into the $S_{1/2}$ ground state, but the detection is deferred until the end of the sequence. The single-qubit Cliffords are global, affecting both ions. Of the slots in the 16 rounds, $n-2$ randomly selected slots contain an entangling gate. We choose $n\in\set{2,10,18}$, which includes one preparation and one readout entangling gate.

An SSB sequence has three possible outcomes: accepted and correct, accepted and wrong, or rejected. The readout gates result in a predetermined ideal outcome of either both ions bright or both dark, and these ideal outcomes are chosen with equal frequency. A shot is rejected if the error checks detect an error. Each error detection check consists of four RF pulses with \SI{729}{\nano\meter} multi-step deshelving, with fluorescence detection delayed until the end of the sequence, as described in \cref{sec:methods-error-detection}. Following error detection, the intended logical state is checked on both ions, with an expected outcome of either both bright or both dark. One or both ions differing from the expected outcome is counted as accepted and wrong.

The reported fidelity from the SSB measurements, $F_\mathrm{SSB}=1-0.19(13)\times 10^{-4}$, is extracted from the fraction of wrong counts $q_{n,t}$ with $n$ entangling gates via $q_{n,t} = 1 - a_t (F_\mathrm{SSB})^{n-2}$. Here, $q_{n,t}$ is the fraction of accepted and wrong counts out of all accepted counts with expected outcome $t$ being either both bright or both dark. The constants $a_t$ capture all effects that do not scale with the number of inserted gates $n-2$, and are fitted separately to account for the different measurement errors of bright and dark states. The fit is performed by binomial maximum likelihood over the three measured values of $n$ and both values of $t$, and the quoted uncertainty is $1\sigma$. We also extract a raw error of 6.3(1)\% by fitting the fraction of trials that were both correct and accepted out of all attempts.
The SSB data were collected over approximately six days under a protocol fixed before data collection. The data include approximately 0.67 million trial attempts of which 0.30 million were accepted. The cumulative estimate of the error per gate, with uncertainty, is shown as a function of the number of trials measured up to that point in \cref{fig:methods-entangling-gate}f. The binomial likelihood gives a one-sided 95\% upper bound of $0.40\times10^{-4}$, and a bootstrap calculation gives an uncertainty of $0.11\times10^{-4}$. The relatively large statistical uncertainty we report (relative to the magnitude of the reported error) is limited by the optimal number of gates at this rejection fraction being relatively small, which does not constrain the error per gate as quickly as the much-longer sequences reported in other experiments. The same set of occasional calibrations were performed as in SLERB (gate time, $\jz^2$ pulse duration, RF frequency and pulse durations, \SI{729}{\nano\meter} frequencies and pulse durations, line-noise feedforward).

The reported infidelity is consistent with our error model's estimate for the error per gate of $0.23(2)\times10^{-4}$, obtained by propagating independently calibrated error sources and their uncertainties through the model (see \cref{sec:methods-budget}). 
The reported SSB infidelity $1-F_\mathrm{SSB}$ is the average over the twelve symmetric stabilizer states prepared in these experiments, but our error model estimates that in this case this is a conservative estimate of the Haar-averaged conditional gate infidelity $1-F_\mathrm{Haar}$. While SSB does not in general provide guarantees on the relationship between these, our detailed error model and simulations of the SSB sequence allow us to estimate $F_\mathrm{Haar}$. Our model predicts that the relationship between the Haar and SSB infidelities under our error sources is $1-F_\mathrm{Haar} \approx 0.72 \times (1-F_\mathrm{SSB})$. We report only the higher-infidelity SSB measurement, not the estimated Haar value, and the difference is within the measurement uncertainty. The experimental data do not provide direct evidence for the gate's action on antisymmetric states.

While SSB attempts to measure only the entangling gate's contribution to the error by holding the number of single-qubit operations and erasure checks constant, it is not entirely possible to disentangle these \cite{Tsa24}. The states that the single-qubit operations act on change with the number of entangling gates, which provides a mechanism for them to influence the measured error per gate. In our error model, imperfections of the amplitude or frequency of the RF pulses in the single-qubit operations contribute $\approx 6\times 10^{-6}$ to the extracted error per gate, about a quarter of the error budget's estimated total. Our reported infidelity is thus a conservative estimate of the error from the entangling spin-dependent force operations themselves.

\subsection{Qudit maximally entangling gate}
This class of entangling gate is also useful beyond logical qubits. Qudit-based computing, where a $d$-dimensional system (with $d>2$) is used within each ion to encode $\log_2(d)$ qubits, has seen a surge of interest recently \cite{Rin21,Hrm22,Low19,Low23,Gre26}. Separate from encoding logical qubits in each ion, it has been suggested that qudit-based encodings may have favorable error correction properties \cite{Cam14,Lim23}, and they reduce the number of atoms needed to store a fixed amount of information. Additionally, for certain circuits, qudit-based encoding can replace some inter-ion entangling interactions with intra-ion single-qudit operations, which can be performed with higher fidelity \cite{Gao22}.

The entangling gate discussed here, which results in an interaction of the form $\exp[-i \theta (\jz \otimes \hat{I} - \hat{I} \otimes \jz)^2 ]$, has interesting properties for qudit-based computing, as it can create a maximally entangled state of two $d$-dimensional systems in a single application of the gate \cite{Oma22}. If both ions are initialized in the state
\begin{align*}
    \Ket{\psi_0} = \sum_{m_1,m_2}\frac{1}{d}\Ket{m_1}\!\Ket{m_2},
\end{align*}
then application of the gate $\exp[-i \pi/d (\jz \otimes \hat{I} - \hat{I} \otimes \jz)^2]$ maps this to a maximally entangled state that is equivalent, up to local $\mathrm{SU}(d)$ unitaries on each ion, to the GHZ-like state
\begin{align*}
    \Ket{\psi_\mathrm{max}} = \sum_m \frac{1}{\sqrt{d}} \Ket{m}\!\Ket{m}.
\end{align*}
For $J=5/2$, that is an entangling gate with $1/3$ the geometric phase of the logical qubit gate, with correspondingly smaller errors from both the shorter gate time and the smaller phase space trajectories that are less sensitive to motional errors \cite{Sut22b}. The resulting entangled state contains $\log_2(d)$ bits of entanglement entropy between the two ions, the maximum possible (\cref{fig:methods-entangling-gate}c). This is true for any spin $J$, not limited to $J=5/2$. Previous qudit gates are either composed of multiple two-level interactions interleaved with single-qudit rotations \cite{Rin21}, or have a limit to the maximum dimension they can entangle fully \cite{Hrm22}, while the $\jz\otimes\jz$ Hamiltonian creates a maximally entangled state in a single shot.

\subsection{Single-qubit gate construction and benchmarking}
\label{sec:methods-1q}

Turning on one of the horizontally polarized Raman beams, which contains equal $\sigma^+$ and $\sigma^-$ components, creates a Hamiltonian proportional to $\jz^2$. The $\mathrm{SU}(2)$ rotations available from the RF field, together with this non-$\mathrm{SU}(2)$ Hamiltonian, are sufficient to generate arbitrary unitaries in a space of arbitrarily high dimension \cite{Oma21,Gio03,Pal03,Cha07b,Hei26}. An arbitrary $\mathrm{SU}(d)$ unitary requires $O(d^2)$ control degrees of freedom. Each RF pulse provides two degrees of freedom and each $\jz^2$ pulse one, and we optimize the pulse parameters numerically so that the full sequence implements the target logical unitary identically on all three shells \cite{Deb23,Shi26}.

\textit{Clifford randomized benchmarking.} Each trial implements a random sequence of single-qubit Clifford gates followed by the Clifford gate that inverts the sequence. For the error-tolerant gate set, we measure the population in both the $\l0$ and $\l1$ states at the end of the experiment to fit both logical errors and leakage into the error subspaces. These two quantities are jointly fit by weighted least squares, leaving the asymptotic value floating to account for population returning from the error subspaces to the logical subspace. For the detection-only gate set, we add an erasure check every 50 gates and stop the sequence if a check flags an error. We fit the erasure, correct, and incorrect counts by multinomial maximum likelihood.

\textit{Injected dephasing errors.} For the added-noise experiments of \cref{fig:clifford-rb}c, the injected $\jz$ error per Clifford is the average error introduced to an ideal logical state by a single application of the random distribution of $\jz$ rotations, given by $5 \sigma^2/4$ where $\sigma$ is the standard deviation of the applied $\jz$ rotation angle distribution in radians. Each error--Clifford pair is repeated 15 times, and results are averaged over many trials, each with a distinct random sequence of Cliffords. The error-detected data come from a variant where the error-correction step is not performed, and it is replaced by an erasure check before readout.

\textit{Logical $T$ gate.} We prepare the six cardinal states in each error shell via a state preparation Clifford acting on $\l0$, $\el0$, and $\eel0$. We then apply the logical $T$ gate followed by a virtual rotation implementing its inverse on the shell being measured and then undo the state preparation Clifford. The virtual rotation is exact and does not undo any of the error of the $T$ gate. The probability of returning to the initial state measures the overlap with the ideal action of the gate. The probability is averaged over the six input states of each shell. For the error-detected result, we discard trials that were detected outside of the logical shell.

\subsection{Error budget}
\label{sec:methods-budget}
\begin{figure*}
    \centering
    \includegraphics[width=1\linewidth]{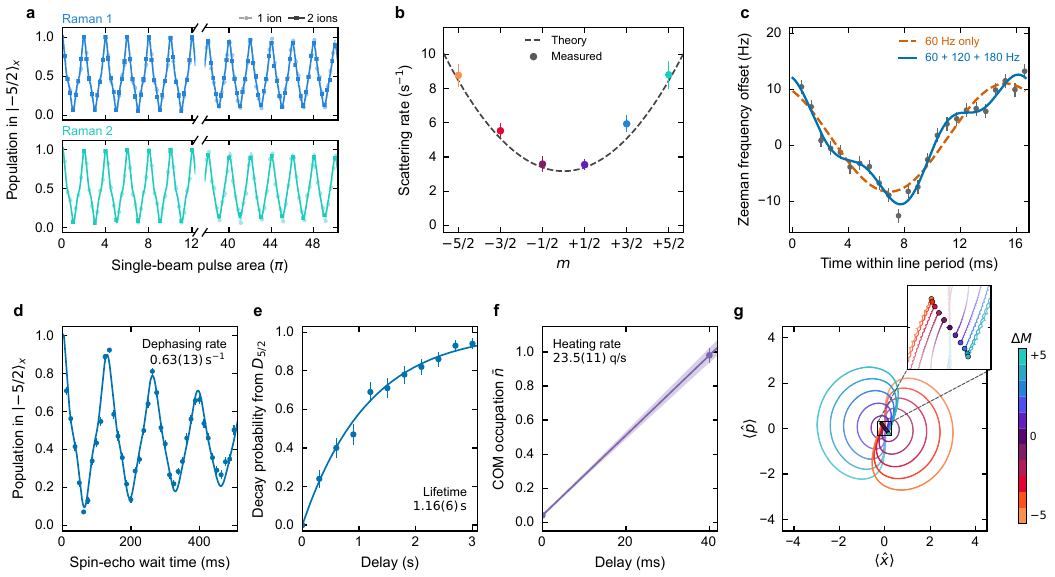}
    \caption{\textbf{Calibrations of the dominant error mechanisms in the logical entangling gate error budget.} 
    \textbf{a,} Light-shift stability measurements of the Raman beam intensity (1 ion) that are also used to extract the amount of unequal illumination (from the difference between 1 and 2 ions).  
    \textbf{b,} Comparison of measured scattering rates to theoretical predictions. Rates are extracted from fits to measurements in each eigenstate of $\jz$ when both Raman beams are turned on at the power used in the entangling gates.
    \textbf{c,} Zeeman frequency shift over one line period, showing the line noise correction applied during the SSB measurements.
    \textbf{d,} Measurement of the background quadrupole shift on the $D_{5/2}$ manifold by preparing an ion in $\ket{-5/2}_X$ and waiting, with a spin echo pulse halfway through the delay. The fit is also used to measure the spin-echo coherence time. 
    \textbf{e,} $D_{5/2}$ lifetime measurement when prepared in the $\ket{-5/2}$ state with no lasers on.
    \textbf{f,} Heating rate measurement of the COM mode after sideband cooling.
    \textbf{g,} Simulated motional phase-space trajectories of each collective projection $\Delta M=m_1-m_2$ for the applied gate waveform, computed to all orders in the Lamb--Dicke parameter, in a frame rotating near the minimum detuning during the gate. Inset: the loop-closure region. Points mark the final displacements.}
    \label{fig:methods-error-budget}
\end{figure*}

We construct an error budget from independent calibrations, with no parameters fitted to the SSB data, for the raw and error-detected SSB error per entangling gate. The error budget is shown in \cref{fig:limits}a and listed explicitly in \cref{tab:supplement-ssb-budget}. We describe each contribution briefly here. A separate error budget for the logical Bell state fidelity is given in the Supplementary Information (\cref{tab:supplement-bell-state-budget}), which includes large state preparation error terms.

These experiments were performed at the settings described in \cref{sec:methods-rb}. Raw error is measured from the decay of the fraction of correct and accepted outcomes as a function of the number of entangling gates, while error-detected (conditional) error is measured from the decay of correct outcomes out of accepted outcomes as a function of the number of entangling gates. All numbers given are per inserted gate. Each contribution is added in first order based on their independent calibrations, leaving out cross terms where different error sources may combine.

The total estimated raw error is $(4.9 \pm 1.3)\%$ and the estimated conditional error is $0.23(2)\times 10^{-4}$, compared to measured raw error of 6.3(1)\% and conditional error of $0.19(13)\times10^{-4}$ per gate. The conditional values are $0.3\sigma$ apart.

For each error term, we state how the probability of error scales under error detection, many of which are plotted in \cref{fig:limits}d and \cref{fig:methods-codewords}d with the vertical line and band representing the estimated raw error in the SSB error budget and its uncertainty. The probability of a raw error, $p$, refers to that mechanism in isolation, and we assume error detection is perfect. Each application of $\jz$ to a logical state moves one step in the $\jx$ basis, and an undetected logical error requires a net $\jz^5$ error to flip the logical state. Similar arguments can be used to derive the fidelity after ideal error correction, where at most second-order $\jz$ errors are correctable in spin-$5/2$.

\textit{Raman pulse-area noise.} Fluctuations in the intensity of either Raman beam, from laser power, beam pointing or polarization, rescale the Hamiltonian as $\hat{H}\mapsto\hat{H}\,[1+\epsilon_\Omega(t)]$, where $\epsilon_\Omega(t)$ is the fractional error of the two-beam Rabi frequency. Because each beam alone also produces a $\jz^2$ tensor light shift on each ion, these fluctuations appear mainly as residual local $\jz^2$ rotations, which the code corrects and detects. The two-beam Rabi frequency is proportional to $|E_1E_2|$, the geometric mean of the single-beam light shifts, each of which is proportional to $|E_i|^2$. Both ions are prepared in the $-5/2$ eigenstate of $\jx$, one Raman beam with Rabi frequency $\Omega$ is turned on for a time $t$ with a spin-echo RF $\pi$ pulse applied midway, and an RF $-\pi/2$ pulse returns the ions, in the absence of any interaction, to the $-5/2$ eigenstate of $\jz$, which is then read out. Under this Hamiltonian, the population that returns to $-5/2$ is
\begin{equation}
\begin{aligned}
    P_{\mathrm{ret}}(\Omega,t)=\frac{1}{256}\bigg[&126+100\cos(6\Omega t)\\
    &+10\cos(12\Omega t)+20\cos(18\Omega t)\bigg]. \label{eq:unequal-f}
\end{aligned}
\end{equation}
We characterize the noise with single-beam light-shift measurements on a single ion, fitting the decay of the oscillations to a Gaussian (shot-to-shot) model (\cref{fig:methods-error-budget}a) and finding pulse-area errors of 0.210(4)\% for beam 1 and 0.223(4)\% for beam 2. When propagated to the pulse areas in the entangling gate, this is the largest contribution to the raw error, at 2.8(1)\%. After detection it is below $10^{-7}$. These intensity fluctuations are $\jz^2$ at lowest order, and a logical flip requires flipping both qubits without leaving the code space, which is $\jz^{10}$ and $O(p^5)$ undetected infidelity.

\textit{Unequal illumination.} To measure the unequal illumination of the two ions, we compare the contrast decay induced by single-beam light-shifts with one ion and with two ions. Any additional contrast decay with two ions is due to unequal illumination of each ion. 
If the ions are illuminated unequally, the two-ion population that returns to $\Ket{-5/2}_X$ based on \cref{eq:unequal-f} is
\begin{align*}
\left[P_{\mathrm{ret}}(\bar{\Omega}_b+\Delta\Omega_b/2,t)
+P_{\mathrm{ret}}(\bar{\Omega}_b-\Delta\Omega_b/2,t)\right]/2,
\end{align*}
where $\bar \Omega_b$ is the average Rabi frequency for beam $b$ and $\Delta \Omega_b$ is the difference in Rabi frequencies of the two ions.
The fits give a fractional half-difference in the intensity seen by the two ions of 0.125(9)\% for beam 1 and 0.141(10)\% for beam 2. This difference affects both the light shifts and the force seen by the ions, and both of those effects are included in the quoted error contribution.
The force ratio is $F_2/F_1=\sqrt{(\Omega_{1,1}/\Omega_{2,1})(\Omega_{1,2}/\Omega_{2,2})}$, where $\Omega_{i,b}$ is the single-beam light shift of beam $b$ on ion $i$ (proportional to its intensity), and we quote an uncertainty that includes lack of knowledge of which beam is brighter on which ion. The effective Hamiltonian is then proportional to $\jz\otimes \hat{I}-(F_2/F_1)\,\hat{I}\otimes\jz$, and the gate implements
\begin{align}
    \hat{U}(F_2/F_1)
    &= \exp\!\Big[
        -i\theta \frac{4}{(1+F_2/F_1)^2}\notag \\
        & \qquad\qquad \times\bigl(\jz\otimes\hat{I}-(F_2/F_1)\hat{I}\otimes\jz\bigr)^2
    \Big].
    \label{eq:unequal-prop}
\end{align}
We apply one loop with $\theta=\pi/4$, a spin-echo $\pi$ pulse, and a second loop with $\theta=\pi/4$. The spin echo maps $\jz\mapsto-\jz$, which leaves this quadratic operator unchanged. In the qubit case, by contrast, $\hat{\sigma}_z^2=\hat{I}$, and unequal illumination can be calibrated away whether or not a spin echo is applied. The relative propagator of the ideal and actual unitaries, to first order in $F_2/F_1-1$, is
\begin{equation}
    \hat{U}(1)^\dagger\hat{U}(F_2/F_1)\approx
    \exp\!\bigl[-i\theta\,(F_2/F_1-1)\bigl(\hat{I}\otimes\jz^2-\jz^2\otimes\hat{I}\bigr)\bigr],
\end{equation}
and the fidelity is $\lvert\braket{\mathbb{00}|\hat{U}(1)^\dagger \hat{U}(F_2/F_1)|\mathbb{00}}\rvert^2$. With the measured parameters, the contribution to the raw infidelity is $1.1 \pm 1.2\%$ and the contribution to the error-detected infidelity is $<10^{-10}$. The large uncertainty range reflects the lack of knowledge of which beam is brighter on which ion, with the error being $2.2\%$ if the same beam is brighter on both and $0.03\%$ if the opposite is true. The error-detected contribution is $<10^{-10}$ in either case. The lowest-order mechanism for a logical error from unequal illumination is two quadratic $\jz^2$ terms on each ion plus a detectable two-ion cross term $\jz\otimes\jz$ acting on both ions, giving an $O(p^6)$ residual error scaling.

\textit{Raman scattering.} The measured rates of Raman scattering out of $D_{5/2}$ are \SI{3.6(3)}{\per\second}, \SI{5.7(3)}{\per\second} and \SI{8.8(5)}{\per\second} for $|m|=1/2$, $3/2$ and $5/2$ (\cref{fig:methods-error-budget}b plots the total scattering rates taking the branching ratio back to $D_{5/2}$ into account, with natural lifetime decay subtracted). The ions are initialized in $\Ket{-5/2}_X$ eigenstate of $\jx$ and the gate does not change spin states. This means that, while the beams are on, the population in each $\jz$ eigenstate $\Ket{m}$ is $\lvert\braket{m \vert -5/2}_X\rvert^2$, so the scattering error is
\begin{align*}
    p_{\mathrm{sc}} = 2t_\mathrm{Raman}\sum_m \Gamma_m \lvert\braket{m \vert -5/2}_X\rvert^2,
\end{align*}
where $\Gamma_m$ is the scattering rate and $\lvert\braket{m \vert -5/2}_X\rvert^2$ is the population of state $\ket{m}$. The effective Raman exposure time $t_\mathrm{Raman}$ is obtained by integrating the beam intensities over both loops and the light-shift compensation pulse.
Most scattering events that leave the $D_{5/2}$ manifold end up in the $D_{3/2}$ or $S_{1/2}$ manifolds and are detected. For the entangling gate's Raman exposure, this gives a scattering probability of $2.06(5)\times10^{-3}$ out of the $D_{5/2}$ manifold, all of which is detected as erasure. About 6\% of scattering events instead return to $D_{5/2}$. After the decoding pulse, some of these events leave the ion in $m=\pm3/2$ or $\pm1/2$, where they are also detected, and the rest are in $m=\pm5/2$, where they are undetectable logical errors. Taking into account the branching ratios to each $\ket m$ state and the dynamics of the decoding pulse, these events contribute $9.9(3)\times10^{-5}$ to the raw error and $1.52(4)\times10^{-5}$ after error detection. Every photon scattered back into $D_{5/2}$ has a chance to flip the logical state, so residual Raman scattering error is $O(p)$.

\textit{Spin dephasing.} We measure residual $\jz$ dephasing by preparing both ions in the $-5/2$ eigenstate of $\jx$ and performing spin-echo Ramsey measurements (\cref{fig:methods-error-budget}d). A small $\jz^2$ term is present even with the lasers off, which we attribute to the electric quadrupole fields that confine the ions \cite{Roo06,Sha16}. We fit the data to a model with coherent $\jz^2$ rotations and white-noise dephasing described by the collapse operator $\sqrt{2\gamma}\,\jz$, for which the off-diagonal elements of the $D_{5/2}$ density matrix decay as $\rho_{m,n}(t)\propto e^{-\gamma(m-n)^2t}$. A fit to multiple such measurements taken during the SSB experiments gives $\gamma=\SI{0.53(6)}{\per\second}$, corresponding to a spin-echo coherence time between neighboring states of $T_{2E}=\SI{1.90(21)}{\second}$. Simulating this dephasing for the \SI{359}{\micro\second} during which the ions are in the $\jx$ eigenstate gives a raw error of $9.4(1.0)\times10^{-4}$, which is fully correctable; after detection it is below $10^{-10}$. Spin dephasing is a $\jz$ process, requiring five steps to flip the logical state and giving an $O(p^5)$ error scaling.

\textit{RF constant detuning error.} While a detuning error is detectable during the application of the spin-dependent force and during delays, a detuning during an RF pulse leads to an undetectable error at second order in the detuning. We estimate the shot-to-shot RF detuning using repeated Ramsey scans. The average detuning error measured in this way gives $1.7(7) \times 10^{-6}$ raw and $1.3(5) \times 10^{-6}$ conditional error per gate. These errors are undetectable, so the scaling is $O(p)$ for all RF detuning errors.

\textit{Line magnetic-field noise.} The \SI{60}{\hertz} line period is visible as a periodic magnetic field shift on the ions, including higher harmonics (\cref{fig:methods-error-budget}c). We trigger experiments on an edge of this signal and compensate for the repeatable portion of this shift by adjusting the phases of RF operations as a function of the elapsed time since the start of the line period \cite{Tat26}. Any residual (uncompensated) field shifts can cause errors at second order (as a $\jx$ rotation with a small $\jz$ detuning error will result in an undetectable $\jy$ error proportional to the square of the detuning). In the SSB experiments reported here, we use feedforward correction to compensate the \SI{60}{\hertz}, \SI{120}{\hertz}, and \SI{180}{\hertz} components. We estimate a residual error of $0.11(10)\times 10^{-5}$ (\SI{60}{\hertz}) and $0.16(10)\times 10^{-5}$ (higher harmonics) from a collection of Ramsey experiments performed with the feedforward active. 

\textit{Differential Zeeman splitting.} The two ions sit approximately \SI{3.8}{\micro\meter} apart, and a spatial gradient in the magnetic field along that direction will result in different Zeeman frequencies for each ion, which will introduce detuning errors that are undetectable at second order. To measure this, we perform Ramsey measurements in the $D_{5/2}$ manifold with both ions simultaneously. We initialize both ions in the $\Ket{-5/2}$ eigenstate of $\jz$ and apply a $e^{-i \jx \pi/2}$ rotation. After an evolution time $t$ we apply a second $\pi/2$ rotation with variable phase $\phi_\mathrm{Ramsey}$. Without any detuning, the average probability of finding the ions in $\Ket{-5/2}$ would be $\cos^{10}(\phi_{\mathrm{Ramsey}}/2)$. With the detunings $f_1$ and $f_2$ of the two ions it becomes $\left[
\cos^{10}\!\left(\frac{\phi_{\mathrm{Ramsey}}-2\pi f_1t}{2}\right)
+\cos^{10}\!\left(\frac{\phi_{\mathrm{Ramsey}}-2\pi f_2t}{2}\right)
\right]/2$, with additional small effects from the quadratic Zeeman shift and magnetic-field dephasing. The magnitude of each ion's detuning from the midpoint extracted from the data is \SI{\pm 1.46(16)}{\hertz}. The RF $\pi$ time is \SI{54.9}{\micro\second}, and so this contributes $\approx 2.8\times10^{-6}$ raw error and $\approx 1.8 \times 10^{-6}$ conditional error.

\textit{RF pulse-area error.} We measure the average over- or under-rotation of RF pulses from sequences of many $\pi$ or $\pi/2$ pulses. The model estimates $0.9(6)\times 10^{-5}$ raw and $8(5)\times 10^{-8}$ conditional error based on these measurements. In our experiments, the encoding pulses and the spin-echo pulses are about orthogonal axes, so a pulse-area error during the spin echo pulse is undetectable and contributes $O(p)$.

\textit{Spin flips from magnetic-field noise.} Near surfaces, atoms experience magnetic-field noise with a frequency spectrum that has measurable components at the Zeeman splitting frequencies \cite{Emm09}. We measure the rate of spin flips by preparing the $\Ket{+1/2}$ state and watching for leakage out of this state into other $D_{5/2}$ states (conditional on no decay to the $S_{1/2}$ ground state). We measure a spin flip rate of \SI{0.013(6)}{\per\second} per ion from the $\Ket{+1/2}$ state, which gives $10(4)\times 10^{-6}$ raw error and $1.6(7) \times 10^{-6}$ conditional error per gate with error detection. We assume that there is not a preferential axis for transverse magnetic field noise. In simulations, about 17\% of spin flips result in undetected errors, giving $O(p)$ scaling with error detection. See the Supplementary Information for more details on this noise source.

\textit{Thermal occupation and effects beyond the Lamb--Dicke approximation.}
The Lamb--Dicke approximation assumes that the force a state sees is independent of its position in phase space. However, for the large trajectories (displacement five times as large as a standard qubit entangling gate) in the logical qubit gate, this approximation breaks down. Out-of-Lamb--Dicke effects cause the trajectories to not return to the origin and the geometric phases to be incorrect (\cref{fig:methods-error-budget}g). To counteract this, for the Bell-state experiments reported in the main text, we employ an adiabatic ramp of the detuning and Rabi frequency based on \cite{Sut23b,Hug25,Bow25}. In the SSB experiments, we optimize for shorter gate times at the expense of less protection from these effects, as that leads to an overall lower conditional error rate. Motional occupation worsens these effects, and we report combined infidelities at the measured COM motional occupation $\bar{n}_\mathrm{COM}=0.041(10)$. The estimated raw infidelity is $4.32(18) \times 10^{-4}$ and the estimated error-detected infidelity is $1.4\times 10^{-9}$. At higher temperatures the error from thermal occupation scales with the width of the distribution and gives an $O(p^5)$ scaling, while at very cold temperatures the probability of error is just proportional to the probability of there being occupation in the first excited state, resulting in an $O(p)$ error. The beyond Lamb--Dicke effects come from noncommuting motional distortions in both loops of the gate, which include a direct quintic phase term giving an $O(p)$ scaling at low error. At higher strengths, higher powers of $p$ become more important, leading to steeper scaling in \cref{fig:methods-codewords}d.

\textit{Motional heating.} Heating of the center-of-mass mode during the gate leaves the phase-space loops unclosed and with incorrect areas. When the motion is traced out, this produces correctable $\jz$-polynomial errors. It contributes $3.37(16)\times10^{-3}$ to the raw error and less than $10^{-10}$ after detection, at the measured heating rate of \SI{23.5(1.1)}{quanta\per\second} of two ions on the COM mode. The heating rate is measured from simultaneous fits to \SI{729}{\nano\meter} blue and red sideband and carrier Rabi oscillations, at delays of 0 to \SI{40}{\milli\second} following sideband cooling (\cref{fig:methods-error-budget}f). Random motional kicks produce phase errors linearly proportional to the collective spin projection $\Delta M$, so the leading term after error detection is $O(p^5)$.

\textit{Mode-frequency variation.} The motional frequency is regularly recalibrated. These calibration measurements give a \SI{28.1(4)}{\hertz} root-mean-square (RMS) spread, which we model as the width of a Gaussian distribution of random static offsets of the motional frequency during the gate. This gives $(3.02 \pm 0.09)\times 10^{-4}$ raw error and $<10^{-10}$ conditional error. An offset in the mode frequency changes both the loop closure and area, and the linear displacement term combines with two quadratic phase terms to give an error-detected scaling of $O(p^3)$.

\textit{Beam-path vibration.} The ions are in a Gifford--McMahon cryogenic vacuum system that experiences vibrations due to the action of the cryocooler. As described in \cite{Plo26}, we used an interferometer where one arm reflects off the coldest cryostat stage to measure time-series displacements of the ions relative to the Raman beams, as well as performing coherence measurements on the two arms of the Raman beams. These measured displacements were fed into simulations of the gate as Doppler shifts on the Raman beam difference frequency and propagated through the SSB experiments. We estimate a raw error of $1.34(14) \times 10^{-3}$ and an error-detected infidelity $<10^{-10}$. These errors behave as mode-frequency errors and have an $O(p^3)$ scaling.

\textit{Spectator modes.} The occupation of other motional modes during the gate can cause a spread of the effective Rabi frequency of the gate through variable Debye--Waller factors \cite{Win97}. The beam geometry means that only the axial stretch mode contributes, which we verified by performing spectroscopy of the radial modes. Using calibrated stretch-mode occupation $\bar{n}_\mathrm{stretch}=0.001(6)$, this gives approximately $0.4^{+3.0}_{-0.4}\times 10^{-6}$ raw error and $<10^{-10}$ error-detected infidelity. The spectator-induced displacements produce quadratic geometric phase errors, so at moderate temperatures the scaling is $O(p^5)$. At very cold spectator-mode temperatures there is a crossover to $O(p)$ where the error is proportional to the probability of the mode being in the first excited state, as in the COM motional occupation case.

\textit{Metastable-state decay.} The probability of decay from $D_{5/2}$ (lifetime $\tau=\SI{1.16(6)}{\second}$) to the ground manifold during the gate is $2t_{\mathrm{dwell}}/\tau$, where $t_\mathrm{dwell} \approx \SI{358.8}{\micro\second}$ is the additional time each ion spends in $D_{5/2}$ per inserted gate (\cref{fig:methods-error-budget}e). This contributes $6.2(3)\times10^{-4}$ raw error, all of which is detected and converted to erasure.

\textit{Missed erasure flags.} We estimate the probability of missing an erasure flag by combining the probability of missing a bright ion during fluorescence detection ($1.8(5) \times 10^{-4}$) with the probability of failing to deshelve error-state population with the \SI{729}{\nano\meter} laser. The final error-detection step uses two fluorescence detection steps in series, giving an estimated conditional error of $8\times 10^{-10}$ per entangling gate.

\subsection*{Data availability}
The data that support the findings of this study are available in a public repository \cite{Deb26}.

\subsection*{Code availability}
The code used for simulation and analysis is available in a public repository \cite{Deb26}.

\section*{Acknowledgments}
DISTRIBUTION STATEMENT A. Approved for public release. Distribution is unlimited.

This material is based upon work supported by the U.S. Army Research Office through grant W911NF-24-1-0379 and by the Under Secretary of War for Research and Engineering under Air Force Contract No. FA8702-15-D-0001 or FA8702-25-D-B002. Support is also acknowledged from the U.S. Department of Energy, Office of Science, National Quantum Information Science Research Centers, Quantum Systems Accelerator (Award No. DE-SCL0000121). ILC acknowledges support in part by the NSF Center for Ultracold Atoms. Any opinions, findings, conclusions or recommendations expressed in this material are those of the author(s) and do not necessarily reflect the views of the Department of Energy and Under Secretary of War for Research and Engineering.

\section*{Author contributions}
KD conceived of and led the experimental work and theoretical analysis, with contributions from AVM. AP, AVM, ERC, DR, WJS, and MEK assisted with the experiment. GJG and JO contributed to the error budget analysis, and XS and TJB contributed to the logical Clifford pulse design. JC, CDB, RM, and ILC contributed to the experimental design and writing of the manuscript.

\section*{Competing interests}
The authors declare no competing interests.

\bibliography{refs}

\clearpage
\section*{Supplementary Information}
\setcounter{figure}{0}
\setcounter{table}{0}
\setcounter{equation}{0}
\renewcommand{\theequation}{S\arabic{equation}}
\renewcommand{\theHequation}{supp.\arabic{equation}}
\renewcommand{\figurename}{Supplementary Fig.}
\renewcommand{\thefigure}{S\arabic{figure}}
\renewcommand{\tablename}{Supplementary Table}
\renewcommand{\thetable}{S\arabic{table}}
\renewcommand{\theHfigure}{supp.\arabic{figure}}
\renewcommand{\theHtable}{supp.\arabic{table}}
\crefalias{figure}{supplementfigure}
\crefalias{figure}{supplementfigure}
\crefalias{table}{supplementtable}
\crefalias{table}{supplementtable}

The Supplementary Information contains full error budget tables for the symmetric stabilizer benchmarking (SSB) sequence (\cref{tab:supplement-ssb-budget}) and the Bell state fidelity measurement (\cref{tab:supplement-bell-state-budget}). It also contains a description of Bell state fidelity terms that are not included in the SSB error budget or that are substantially different. We also give detailed timing information about the experimental sequence and measurements of the ion spacing.

\subsection{Experimental sequence}
\begin{figure*}
    \centering
    \includegraphics[width=\linewidth]{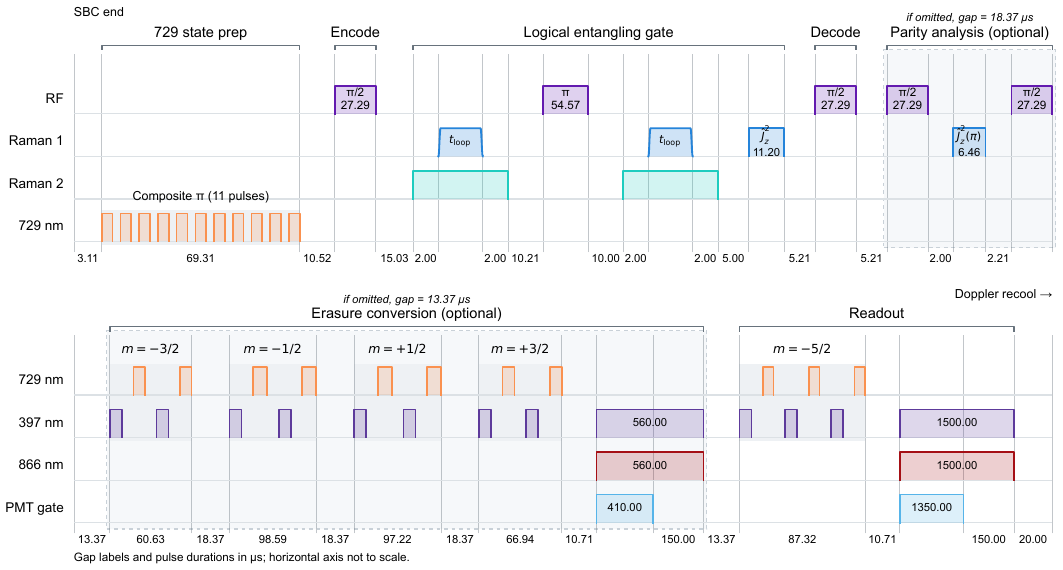}
    \caption{\textbf{Logical Bell-state pulse sequence.} Detailed pulse sequence corresponding to the partial state tomography measurements in \cref{fig:logical-gate}a. Top: preparation of $\Ket{-5/2}$ with a composite \SI{729}{\nano\meter} pulse; encoding; two gate loops of duration $t_{\mathrm{loop}}$, where Raman 1 is shaped and Raman 2 is switched on and off, separated by an RF spin-echo $\pi$ pulse and followed by the $\jz^2$ correction pulse; decoding; parity-analysis pulses, which are applied only in the parity scan. Bottom: error detection (erasure conversion) step, in which population in $\Ket{\pm1/2}$ and $\Ket{\pm3/2}$ is deshelved by \SI{729}{\nano\meter} pulses interleaved with \SI{397}{\nano\meter} pulses and detected by fluorescence; readout of $\Ket{-5/2}$ if the error detection step did not detect an error. Pulse durations and gaps are in \si{\micro\second} and the horizontal axis is not to scale.}
    \label{fig:sequence}
\end{figure*}

\Cref{fig:sequence} shows the timed sequence of one trial of the error-detected Bell-state experiment.

\subsection{Raman two-photon error recovery}
\begin{figure}
    \centering
    \includegraphics[width=1\linewidth]{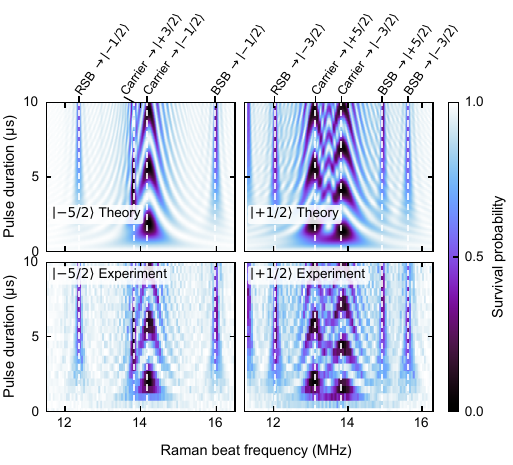}
    \caption{Single-ion Raman spectroscopy showing how the $\jz^2$ light shifts from the Raman beams can create spectral separation of the otherwise-degenerate two-photon transitions. Top row is theory with no free parameters, bottom row is experimental data. Left column is initialized in $\Ket{-5/2}$ which only has one possible transition, while right column is initialized in $\Ket{+1/2}$ which can couple to either $\Ket{-3/2}$ or $\Ket{+5/2}$.}
    \label{fig:supplement-raman-internal-transitions}
\end{figure}

In the experiments in this work, we use the Raman beams to drive motional sideband transitions to apply the error recovery operator. In our previous work \cite{Deb25}, we used $\SI{729}{\nano\meter}$ light to apply carrier and sideband pulses that performed the measurement-free error recovery sequence, but the coherence and light shifts of the \SI{729}{\nano\meter} degraded performance. The Raman beams used here are each polarized so that they have equal $\sigma^\pm$ components and minimal $\pi$ component. This means that in addition to the spin-dependent forces, they can also drive internal-state $\Delta m=\pm2$ transitions and motional sidebands, and the effective linewidth is very narrow. The problem, however, is that with a linear Zeeman splitting, trying to drive transitions from $\ket{-1/2}\rightarrow\ket{-5/2}$ will also drive $\ket{-1/2}\rightarrow \ket{+3/2}$, as both transitions have the same frequency. A similar example is shown in the right column of \cref{fig:supplement-raman-internal-transitions}, where there are carrier terms from $\Ket{+1/2}$ to both $\Ket{-3/2}$ and $\Ket{+5/2}$ that overlap with each other and interfere. Even in the left column, the four-photon process connecting to a state $\Delta m=4$ away is visible. These degeneracies are broken by the $\jz^2$ light shift that the two Raman beams each create individually, which results in the frequency shifts between the carrier features. The top row of \cref{fig:supplement-raman-internal-transitions} is a numerical prediction of the Rabi lineshapes using only independently measured values of the Zeeman splitting, the motional frequency, and the intensity of each beam on the ion. The plot only shows one ion, but with two the COM and stretch modes are both present, enabling the two-ion recovery described in the Methods.

We enhance this spectral separation in three ways in order to drive only the desired operations during the error recovery sequence. First, we drive sideband transitions that are narrower than the carrier transitions by a factor of the Lamb--Dicke parameter. Second, we decrease the power in one of the Raman beams (\cref{fig:logical-gate}d), which decreases the two-photon Rabi frequency more rapidly than it decreases the total $\jz^2$ light shift (the light shift from the other beam is unaffected). Third, we ramp the amplitude of the pulse with a $\sin^2$ profile over several $\si{\micro\second}$ to decrease the spectral width of the pulses. These three techniques lead to negligible coupling to unwanted transitions when driving the sideband pulses.

\subsection{Symmetric stabilizer benchmarking error budget}
\begin{table*}[htbp]
\centering
\caption{\textbf{Logical entangling-gate error budget for symmetric stabilizer benchmarking.} Estimated raw and undetected error after error detection per inserted entangling gate. The contribution from each listed mechanism is estimated from independent calibrations with no parameter fitted to the SSB data (Methods and Supplementary Information). Uncertainties are propagated uncertainties of the calibration inputs and are added in quadrature, except for scattering rows which share a calibration dataset are added linearly. The total predicted value is a first-order sum of the individual mechanisms. This model does not capture all effects of multiple combined errors. The measured raw error is extracted from the fit of the accepted and correct probability as a function of the number of gates.}
\label{tab:supplement-ssb-budget}
\begin{tabular}{lcc}
\toprule
Error source
 & Uncorrected ($\times 10^{-4}$)
 & Error detection ($\times 10^{-4}$) \\
\midrule
Scattering returns to $D_{5/2}$
 & $0.985 \pm 0.025$ & $0.152 \pm 0.004$ \\
\SI{60}{\hertz} magnetic field
 & $0.68 \pm 0.03$ & $0.011 \pm 0.010$ \\
\SI{120}{\hertz} and \SI{180}{\hertz} magnetic field
 & $0.018 \pm 0.012$ & $0.016 \pm 0.010$ \\
Ion--ion Zeeman splitting
 & $0.028 \pm 0.006$ & $0.018 \pm 0.004$ \\
RF over-rotation
 & $0.09 \pm 0.06$ & $<0.001$ \\
RF common detuning
 & $0.017 \pm 0.007$ & $0.013 \pm 0.005$ \\
Transverse field noise (spin flips)
 & $0.10 \pm 0.04$ & $0.016 \pm 0.007$ \\
Missed erasure flags
 & --- & $<0.001$ \\
\addlinespace
Raman intensity noise
 & $281 \pm 11$ & $<0.001$ \\
Unequal beam illumination
 & $112 \pm 125$ & $<0.001$ \\
Raman scattering out of $D_{5/2}$
 & $20.6 \pm 0.5$ & $0$ \\
COM thermal occupation and beyond-Lamb--Dicke effects
 & $4.32 \pm 0.18$ & $<0.001$ \\
COM motional heating
 & $33.7 \pm 1.6$ & $<0.001$ \\
Displacement noise (beam-path vibration)
 & $13.4 \pm 1.4$ & $<0.001$ \\
Mode-frequency drift
 & $3.02 \pm 0.09$ & $<0.001$ \\
Spin dephasing
 & $9.4 \pm 1.0$ & $<0.001$ \\
$D_{5/2}$ decay during gate
 & $6.20 \pm 0.30$ & $0$ \\
Axial spectator Debye--Waller noise
 & $0.004_{-0.004}^{+0.030}$ & $<0.001$ \\
\midrule
Predicted total
 & $485 \pm 125$ & $0.23 \pm 0.02$ \\
Measured SSB error per gate
 & $632 \pm 11$ & $0.19 \pm 0.13$ \\
\bottomrule
\end{tabular}
\end{table*}

\subsection{Logical Bell-state error budget}

We estimate the Bell-state error from independent calibrations propagated to the logical Bell state through our error model. In contrast to the randomized benchmarking experiments in the main text, a single Bell state preparation is sensitive to state preparation and measurement errors. For each mechanism we compute the raw error, the error that remains after ideal local recovery, and the undetected error that remains after ideal detection. The predicted total is a first-order sum and does not capture compounding errors. The predicted raw fidelity of 91(3)\% compares to a measured value of 92(1)\%, and the predicted error-detected fidelity of 99.84(4)\% compares to the measured $99.8^{+0.1}_{-0.2}\%$. The individual terms are listed in \cref{tab:supplement-bell-state-budget}.

\begin{table*}[htbp]
\centering
\caption{\textbf{Logical Bell-state error budget.} Estimated raw error, error after ideal local recovery, and undetected error after ideal detection for each mechanism, each estimated from independent calibrations (Methods and Supplementary Information). We estimate the effect of each error source separately and sum them to obtain the total error. The predicted total is thus a first-order sum of the individual mechanisms and does not capture compounding errors, but these are expected to be small compared to the dominant light shift and SPAM error terms. The ideal correction column assumes perfect recovery of $\jz$ and $\jz^2$ errors on both ions, and the ideal detection column assumes no missed erasure flags. The uncertainties are added linearly to obtain the total uncertainty.}
\label{tab:supplement-bell-state-budget}
\begin{tabular}{lccc}
\toprule
Error source
 & Uncorrected ($\times 10^{-4}$)
 & Ideal correction ($\times 10^{-4}$)
 & Error detection ($\times 10^{-4}$) \\
\midrule
729 nm preparation and readout
 & $54 \pm 8$ & $54 \pm 8$ & $15 \pm 4$ \\
Raman $\hat J_z^2$ pulse-area noise
 & $500 \pm 30$ & $18.8 \pm 2.4$ & $<0.001$ \\
Spin dephasing
 & $5.5 \pm 0.7$ & $<0.001$ & $<0.001$ \\
$D_{5/2}$ decay during gate
 & $4.8 \pm 0.2$ & $4.8 \pm 0.2$ & $0$ \\
Transverse magnetic-field noise (spin flips)
 & $0.059 \pm 0.025$ & $0.053 \pm 0.023$ & $0.014 \pm 0.006$ \\
COM motional heating
 & $12.1 \pm 0.4$ & $<0.001$ & $<0.001$ \\
Axial spectator Debye--Waller noise
 & $0.10 \pm 0.04$ & $<0.001$ & $<0.001$ \\
Beyond-Lamb--Dicke effects
 & $4.65$ & $2.43$ & $<0.001$ \\
Raman scattering out of $D_{5/2}$
 & $67.3 \pm 1.7$ & $67.3 \pm 1.7$ & $0$ \\
Scattering returns to $D_{5/2}$
 & $3.26 \pm 0.07$ & $2.11 \pm 0.05$ & $0.588 \pm 0.013$ \\
Unequal beam illumination
& $200_{-154}^{+282}$ & $1.3_{-1.3}^{+4.2}$ & $<0.001$ \\
Displacement noise (beam-path vibration)
 & $1.69 \pm 0.17$ & $<0.001$ & $<0.001$ \\
Mode-frequency drift
 & $0.08 \pm 0.03$ & $<0.001$ & $<0.001$ \\
RF-detuning preparation/readout error
 & $0.876 \pm 0.016$ & $0.738 \pm 0.013$ & $<0.001$ \\
\midrule
Predicted total
 & $855_{-195}^{+324}$ & $151 \pm 15$ & $15.7 \pm 3.8$ \\
Measured Bell-state error
 & $785 \pm 110$ & --- & $24.5_{-8.2}^{+23.8}$ \\
\bottomrule
\end{tabular}
\end{table*}

\subsubsection{Optical state preparation and readout}
State preparation and measurement on the \SI{729}{\nano\meter} transition were calibrated from two-ion preparation and shelving trials. Preparation failures, which occur with probability 0.39(7)\%, are detected by a fluorescence check before deshelving. This contributes 0.54(8)\% to the raw and recovered errors, including both state preparation failures and readout failures. Under error detection the flagged preparation failures are rejected, leaving a readout assignment error of 0.15(4)\% on accepted trials.

\subsubsection{RF state preparation and readout}
Imperfect RF pulses similarly affect the fidelities, primarily from a magnetic field gradient that causes the two ions' resonances to be different from each other. This gradient was substantially larger in the Bell state measurements than in the SSB measurements with a difference frequency of \SI{36.2(3)}{\hertz}. To calculate the resulting error, we numerically simulate the RF rotations on both ions, with one ion detuned from resonance by \SI{+18.1(2)}{\hertz} and the other by \SI{-18.1(2)}{\hertz}, and a $\pi$ pulse duration of \SI{55.4}{\micro\second}. This gives a raw error of $0.876(16)\times10^{-4}$. With ideal recovery the error is $0.738(13)\times 10^{-4}$. After post-selecting against population detected in the $m=\pm3/2$ and $\pm1/2$ states, the undetected error is $<10^{-8}$.

\subsubsection{Intensity noise and unequal illumination}
We performed similar experiments for the Bell-state fidelity experiments as the unequal illumination experiments shown in \cref{fig:methods-error-budget}a, evaluating both the differential light shift and force imbalance contributions. The fits give single-ion intensity fluctuations of 0.34(1)\% and 0.24(1)\% and a fractional difference in the intensity seen by the two ions of 0.47(4)\% for beam 1 and 0.15(9)\% for beam 2.
Unequal intensities change the force and $\jz^2$ rotation on each ion. For the measured illumination differences this gives $2.0^{+2.8}_{-1.5}\%$ raw error, $1.3^{+4.2}_{-1.3}\times10^{-4}$ error with ideal correction, and $<10^{-8}$ error with detection.

\subsubsection{Effects beyond the Lamb--Dicke approximation}
To counteract the motional loop closure and geometric phase errors caused by the breakdown of the Lamb--Dicke approximation we employ an adiabatic ramp of the detuning and Rabi frequency based on \cite{Sut23b,Hug25,Bow25}. For the Bell state measurements we used a substantially more adiabatic frequency profile compared to the SSB measurements. This gives a raw infidelity of $4.65\times 10^{-4}$, the infidelity after ideal error correction to $2.43\times 10^{-4}$, and the ideal error-detected infidelity to $8.6\times 10^{-10}$.

\subsubsection{Spectator modes}
Debye--Waller factors describe the effect of nonzero motional occupation of the spectator modes on the Rabi frequency of the gate mode. If spectator mode $i$, with Lamb--Dicke parameter $\eta_i$ (which includes the angle between the mode direction and the effective wavevector), is in Fock state $n_i$, the Rabi frequency is scaled as
\begin{align*}
    \Omega=\Omega_0e^{-\eta_i^2/2}L_{n_i}(\eta_i^2)\approx\Omega_0\left[1-\eta_i^2\left(n_i+\tfrac{1}{2}\right)\right]\equiv\Omega_0D_i(n_i),
\end{align*}
where $\Omega_0$ is the Rabi frequency without spectator modes and $L_n$ is the $n$th Laguerre polynomial \cite{Win97}. If the effective wavevector is exactly parallel to the trap axis, the Lamb--Dicke parameters of all modes except the axial center-of-mass and stretch modes vanish, and only one Debye--Waller factor matters. Using the measured stretch mode excitation of $\bar{n}_\mathrm{stretch} = 0.020(8)$ and stretch mode Lamb--Dicke parameter of $\eta_\mathrm{stretch}\approx0.047$, we estimate a raw infidelity of $1.0(4)\times 10^{-5}$, and detected and corrected infidelities both $<10^{-8}$. We note the importance of cooling the spectator modes, however. If the stretch mode were only Doppler cooled to $\bar{n}=5.4(5)$, then the raw error would be $1.6(2)\%$, with a corrected error of $8(3)\times10^{-4}$ and a detected error of $\approx 1\times 10^{-5}$.

\subsubsection{Polarization impurity}
\begin{figure}[htb]
    \centering
    \includegraphics[width=1\linewidth]{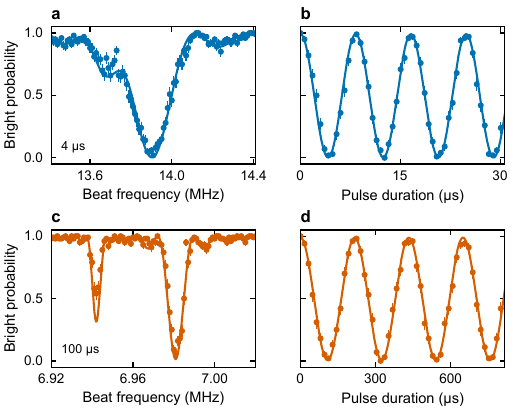}
    \caption{Rabi frequencies of $|\Delta m|=1$ and $|\Delta m|=2$ Raman transitions within $D_{5/2}$, used to bound the vertically polarized component of the Raman beams. \textbf{a,b}, Spectroscopy and Rabi oscillations of the $|\Delta m|=2$ transition. \textbf{c,d}, The same for the $|\Delta m|=1$ transition. The curves in \textbf{b,d} are sinusoidal fits, while the curves in \textbf{a,c} are simulated spectra based on independent measurements of the beam intensities and the Rabi frequency values measured in \textbf{b,d}. In \textbf{c}, it is assumed that each beam has equal $\pi$ polarization, which was not independently measured.}
    \label{fig:polarization-ratio}
\end{figure}
The gate relies on purely horizontally polarized light, whose $\sigma^+$ and $\sigma^-$ components form the polarization gradient that drives the $\jz\otimes\jz$ interaction. A vertically polarized component in both beams would also interfere and produce an effective $\jz^2\otimes\jz^2$ interaction. This is an entangling interaction that cannot be removed with local operations or spin-echo pulses, and it would contribute to the raw infidelity but is detectable at leading order.

We measure the horizontally and vertically polarized powers $P_H$ and $P_V$ by driving Raman transitions between levels of the $D_{5/2}$ manifold. Purely horizontal light drives only $|\Delta m|=2$ transitions, whereas a vertical component in at least one beam also allows $|\Delta m|=1$ transitions. These occur at half the frequency difference and are therefore resolved spectrally (\cref{fig:polarization-ratio}). The fitted ratio of the Rabi frequencies is 26.17(5). Taking into account the Clebsch--Gordan coefficients, the vertical polarization extinction ratio is at least \SI{37.4}{\decibel}, or $P_V/P_H\le1.83\times10^{-4}$. We quote a bound because of the complexity of the multilevel dynamics and because we assume the vertical component is equal in both beams. This tensor force is proportional to the product of the vertically polarized intensities of both beams.

The vector and tensor forces drive the same motional mode, however, so there can be effects that go as the product of vertical and horizontal polarizations, which would give an effect linear in each beam's vertical component. Any odd terms are reversed by the spin echo pulse, leaving only tensor terms. This can be bounded by $\epsilon_Q\le(525\pi^2/64)(P_V/P_H)^4\le10^{-13}$, and we estimate that its contribution to the raw infidelity is negligible.

The geometric phase scales as the square of the effective Rabi frequency, which scales with the beam power. The unwanted angle of the quadrupolar interaction therefore scales as $\theta_Q\sim(P_V/P_H)^2$, and the associated error as $\epsilon_Q\sim(P_V/P_H)^4$ for small $\theta_Q$. Vertical polarization contributes negligibly to the gate error.

\subsubsection{Spin flips}
\begin{figure}
    \centering
    \includegraphics[width=1\linewidth]{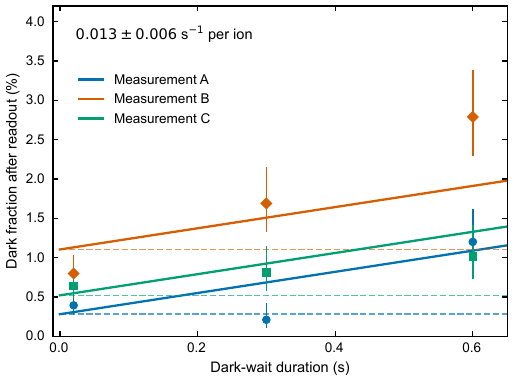}
    \caption{Measurements of the rate of spin flips in the $D_{5/2}$ manifolds when prepared into the $\Ket{+1/2}$ state. The three measurements were taken several hours apart, and the quoted uncertainty takes into account the apparent drift between the measurements.}
    \label{fig:supplement-spin-flip}
\end{figure}
The codes used here protect against $\jz$ errors. Error detection also provides protection from $\jy$ errors. However, a $\jx$ error is a logical phase error and is undetectable. A process that causes within-$D_{5/2}$ spin flips from one magnetic sublevel to another can result in $\jx$ errors. Such processes include magnetic field noise at the Zeeman splitting frequency, such as from the surface trap \SI{50}{\micro\meter} from the ion \cite{Emm09}. To estimate this effect, we prepare the state $\Ket{+1/2}$ (which has strong matrix elements under $\jx$) and wait different durations before measuring the ion's state. We first look for and discard any population that has decayed to the $S_{1/2}$ manifold, and then measure population in the $\Ket{+1/2}$ state. Any dark population has left $\Ket{+1/2}$ due to a spin flip. We measure a spin flip rate of \SI{0.013(6)}{\per\second} per ion (\cref{fig:supplement-spin-flip}), corresponding to an error per Bell state entangling gate of $1.4(6)\times10^{-6}$ after ideal error detection. This measurement corresponds to a transverse magnetic-noise amplitude of \SI{0.75(16)}{\pico\tesla\per\sqrt{\hertz}}, which is comparable to or slightly less than the values measured in \cite{Emm09} when adjusting for temperature and surface-electrode distance (\SI{6}{\kelvin} and \SI{50}{\micro\meter} in our case). Our trap is aluminum, compared to a gold surface in \cite{Emm09}.

\subsubsection{Scattering}
The Bell state experiments use a \SI{1.776}{\tera\hertz} red detuning of the Raman beams from the \SI{854}{\nano\meter} transition, compared to a \SI{5.15}{\tera\hertz} red detuning in the randomized benchmarking experiments. This increases the total scattering probability to $7.2(2)\times 10^{-3}$ per gate, leading to the values listed in \cref{tab:supplement-bell-state-budget} for the scattering outside of $D_{5/2}$ and within $D_{5/2}$.

\subsubsection{Dephasing}
The decoherence rate measured at the time of the Bell state fidelity measurements was \SI{0.50(6)}{\per\second} with spin echo, giving an estimated raw error of $5.5(7)\times 10^{-4}$. With both ideal correction or detection it is $<10^{-8}$.

\subsection{Ion spacing}
The geometric phase gate applies a spin-dependent force to the two ions, causing the collective spin states to be displaced in phase space at a rate proportional to the sum of the forces on each ion's spin state. The spacing of the two ions, relative to the effective Raman wavelength, determines the relative phase between the forces on each ion. If the Raman beam difference frequency is set near the COM mode frequency, as in these experiments, the ion spacing should be set to either an integer number of effective Raman wavelengths (producing a $\jz\otimes\jz$ interaction) or a half-integer number of wavelengths (producing a $-\jz\otimes\jz$ interaction).

\begin{figure}[htb]
    \centering
    \includegraphics[width=\linewidth]{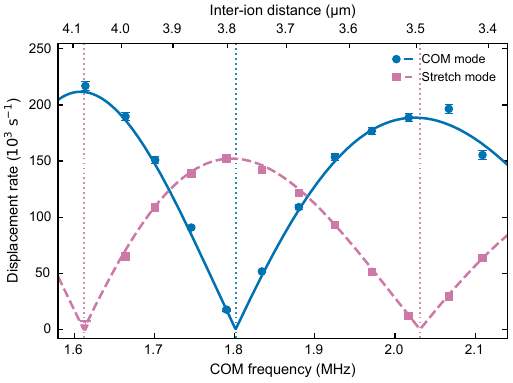}
    \caption{Measurement of the displacement rate as a function of axial trap frequency for the COM and stretch modes when the ions are prepared in the symmetric state $\Ket{-5/2}\!\Ket{-5/2}$. The fits to displacement rate as a function of ion spacing enable a measurement of the effective wavelength of the Raman beams projected along the trap axis. The top axis (inter-ion distance) is calculated from the measured COM frequency $f_\mathrm{COM}$ using the mass and charge of $^{40}\text{Ca}^+$ ions to be $\approx \SI{5.61}{\micro\meter} \times (\SI{1}{\mega\hertz}/f_\mathrm{COM})^{2/3}$. }
    \label{fig:spacing}
\end{figure}

We measure the ion spacing relative to the Raman wavevector by adjusting the axial trap frequency \cite{Win97}. At each trap frequency, we measure the displacement of $\ket{-5/2}\!\ket{-5/2}$ eigenstates of two ions under resonant state-dependent drives at the COM and stretch frequencies. We measure the internal state coherence after a displacement pulse as a function of the pulse duration and fit the decay of coherence as a function of pulse time to give a displacement rate. The measured displacement rates are shown in \cref{fig:spacing}. We then fit the displacement rates to extract the effective Raman wavelength. This is slightly shorter than the expected effective wavelength, likely due to imperfect alignment of the two Raman laser beams relative to each other. An angle between the beams of approximately $95^\circ$, as opposed to the intended $90^\circ$, would lead to this effect.

For the results reported in this manuscript, the trap frequency is approximately \SI{1.801}{\mega\hertz}. The resulting ion spacing results in the COM displacement rate going to zero for symmetric spin states but maximizes the displacement rate of opposite-spin states, leading to a $-\jz\otimes\jz$ interaction. This is because the COM drive displaces states with differing magnetic quantum numbers $m_1$ and $m_2$, while states where $m_1=m_2$ have equal and opposite forces and are not driven. If the ion spacing were such that the COM and stretch displacement rates were equal for equal-projection states (meaning that symmetric and antisymmetric states are equally driven on the COM mode) no entangling gate would be possible.

\end{document}